\documentclass[aps,pra,reprint,superscriptaddress,floatfix]{revtex4-2}
\usepackage[utf8]{inputenc}
\usepackage{amsmath,amssymb,amsfonts}

\usepackage{bm}
\usepackage{graphicx} 
\usepackage{physics}
\usepackage[monochrome]{xcolor}
\usepackage{bbold}
\usepackage{hyperref}

\begin{document}

\title{Collision models for open quantum systems coupled to finite environments}

\author{Gyaneswar Bhoi}
\email{gyaneswarbhoi@iisertvm.ac.in}
\affiliation{School of Physics, Indian Institute of Science Education and Research, Thiruvananthapuram, India-69551}

\author{Anil Shaji}
\email{shaji@iisertvm.ac.in}
\affiliation{School of Physics, Indian Institute of Science Education and Research, Thiruvananthapuram, India-69551}

\date{\today}
\begin{abstract}
We study a system qubit repeatedly interacting with the same environmental qubit, with a reservoir acting on the environment between collisions via a completely positive, trace-preserving map. We show that complete suppression of system--environment correlations uniquely requires a full environmental reset, recovering a semigroup dynamics with a time-independent Gorini--Kossakowski--Sudarshan--Lindblad generator, whereas a partial reset yields a continuous transition between Markovian and non-Markovian regimes governed by a single dimensionless relaxation parameter. For a resonant excitation-exchange interaction, we obtain exact closed-form expressions for the Bloch-vector dynamics for both a generalised depolarising channel and a generalised amplitude-damping channel acting as the reservoir-induced map. Using the Breuer--Laine--Piilo measure and a Choi-matrix CP-divisibility witness, we identify three distinct dynamical regimes across the parameter space: CP-divisible Markovian dynamics, CP-indivisible but P-divisible dynamics, and non-P-divisible non-Markovian dynamics. The boundaries between these regimes, and the structural differences between uniform and anisotropic environmental relaxation, are characterised numerically.

\end{abstract}

\maketitle

\section{Introduction}

An interesting but not very widely appreciated transformation is happening in quantum technologies with respect to how the environment of a quantum system is viewed and understood. While it is understood that 
realistic quantum systems cannot be perfectly isolated from their surroundings traditionally, the environment was treated as an adversary that inevitably produces relaxation, dissipation, and decoherence~\cite{1Breuer-book, 2Rivas-OQS-review, 3Weiss2008quantum-dissipative-system, 4Davies1976-Open-quantum-book, 5Zurek-Decoeherence-Classicality, 6Zurek-Decoeherence-Einseletion, 7Mschlosshauer2007decoherence, 8Preskill2018quantumcomputinginNISQ, 19All-open-dynamics}. The experimental challenge in this case was to isolate the quantum system as much as possible from the environment. Implicit in this approach was often the assumption that the environment of the quantum system is a much larger system with a large number of uncontrollable degrees of freedom. 

This picture has been progressively changed by the increasing ability to not only fabricate, characterize and control individual quantum systems but also the ability to  wield a certain degree of control on the nature and dynamics of the immediate environment that the quantum system of interest interacts with. In other words, it is increasingly possible to not only engineer quantum systems but also to engineer its interaction with its environment and even to engineer the environment itself~\cite{Myatt2000, Turchette2000, Leghtas2015, KimchiSchwartz2016, Liu2024dissipation, GoogleQAI2023}.  A comprehensive survey of the experimental landscape and the underlying theoretical framework of quantum reservoir engineering is given in~\cite{Harrington2022}. The effective size of the Hilbert space of the environment that interacts with the system has also come down with these advances. The environment is no longer an anonymous thermodynamic bath with infinitely many modes and very small correlation times.  It is a finite and often  structured, quantum-mechanical object like an auxiliary transmon qubit, a readout cavity mode, a handful of two-level charge fluctuators in a Josephson junction~\cite{Mueller2019}, or a residual nuclear spin bath in isotopically enriched silicon~\cite{Yoneda2018,Veldhorst2014}.  In each of these engineered settings, the environment possesses its own internal coherence timescale, its own quantum state, and its own dynamics and these properties change in response to what the system does.  The environment therefore is endowed with a memory and the open dynamics of the system becomes non-Markovian.

A small environment endowed with memory requires for its description a mathematical framework that is more general and more flexible than the established approaches. The effective evolution of any initial state of the subsystem of interest, irrespective of the nature of the environment, is obtained by tracing out the environmental degrees of freedom. However, in order to have a useful description of the dynamics of the system, a dynamical map that is well defined on at least a dense subset of the state space of the system, if not the entire state space, has to be constructed~\cite{24ECG-stochastic-PhysRev.121.920,25Krauss-state-effect-operation-1983}. A completely positive, trace preserving (CPTP) map can be constructed if the system and its environment are initially in a product state of the form $\rho_{\rm SE}(0) = \rho_{\rm S}(0)\otimes\rho_{\rm E}(0)$. This assumption allows the
{\em assignment} $\rho_{\rm S}(0)\mapsto\rho_{\rm S}(0)\otimes\rho_{\rm E}(0)$ to be made unambiguously for \emph{any} input state $\rho_{\rm S}(0)$, yielding a map that is linear on the full state space and CPTP. Such a map can be written in the form, 
\begin{equation}
    \Phi^t[\rho] = \sum_k K_k(t)\,\rho\,K_k^\dagger(t), \qquad  \sum_k K_k^\dagger(t)\,K_k(t) = \mathbb{I},
    \label{eq:Kraus2}
\end{equation}
If the initial system-environment state is correlated — as is generally the case for small environments  — the dynamical map is generically not CPTP, and an unambiguous assignment map is not available either.

Before focusing on a small environment, we remark that if the dynamical map is CPTP and it is also divisible, then one can obtain an equivalent description of the dynamics of the open system in terms of a  Gorini--Kossakowski--Sudarshan--Lindblad (GKSL) master equation.~\cite{1Breuer-book,2Rivas-OQS-review, 3Weiss2008quantum-dissipative-system, 4Davies1976-Open-quantum-book, 17GKSL-Form_10.1063/1.522979, 18Lindblad-Form-1976, 27Nakazima-equation, 28Zwanzig-Ensemble-method, 19All-open-dynamics, 10NielsenChuangQuantum, 25Krauss-state-effect-operation-1983, 26M.M.WOLF-quantum-channels-and-operations-guiede-tour, 23CHOI1975285}. Effectively, obtaining the GKSL master equation requires not only that the joint system-environment state remains a product state at all times but that the environment state itself is stationary in time. Assuming a large environment which is in an equilibrium state with very short relaxation times and weak coupling to the system, so that changes in the state of the environment induced by the dynamics of the system are not only small in magnitude but the environment returns to equilibrium rapidly, naturally leads to evolution that is captured accurately by a suitable GKSL master equation. Formally, these assumptions are termed the Born and Markov approximations, respectively.  When these assumptions are not applicable, the open dynamics of the system becomes non-Markovian, characterized by the back-reaction of the environment on the system, which can be due to finite environments, strong coupling, the buildup of system-environment correlations or a combination of all these factors~\cite{30Chruscinski-Markovian-criteria_2012, 31Chruscinski-divisibilty-Vs-backflow-PhysRevA.83.052128,19All-open-dynamics}. 

Once the environment is small enough that both classical and quantum correlations can build between the system and the environment, even the existence of a dynamical map is not guaranteed from which it follows that a master equation description is also not available. In the following we focus on infinitesimal- or finite-time dynamical maps rather than on master equations. The conceptual, technical and mathematical challenges of constructing dynamical maps in the presence of correlations between the system and its environment and their possible resolutions have been discussed extensively in the literature~\cite{Pechukas1994, Alicki1995, Pechukas1995, Stelmachovic2001, JordanShajiSudarshan2004}. We will not dwell on these issues, but rather will intentionally sidestep these issues by using collision models to construct the dynamical maps of interest. 

Collision models provide a versatile microscopic framework for studying open-system dynamics and memory effects, with non-Markovian effects introduced through multiple mechanisms like ancilla--ancilla interactions, repeated contacts, or pre-correlated states~\cite{35J_Rau_1963_PhysRev.129.1880, 36scarani2002thermalizingCM, 37ziman2005descriptionCM, 38Ciccarello2017_quantumOptics, 39Vachini2021collisionModel, 40CM_A-beginner-guidecusumano2022quantum, 41CICCARELLO20221, 42Brun-quantumtrajectory-CM, 43Ziman-description-of-OQD-CM, 44Ciccarello-CM-to-non-Markovian-PhysRevA.87.040103, 45Ciccarello-CM-non-Markovian-Incoherent-Swap_2013, 46Non-markovianity-system-env-correlation-CM-PhysRevA.89.052120, 47CM-All-optical-non-markovian-PhysRevA.91.012122, 48CM-class-of-exact-memory-kernel-ME-PhysRevA.93.052111, 49CM-for-non-Markovian-dynamics-PhysRevA.94.012106, 50CM-Non-Markovianity-and-coherence-and-system-environment-correlations-in-a-long-range-PhysRevA.96.022109, 51Cm-Rybar-indivisibleCM, 52CM-environmental-correlation-PhysRevA.90.032111, 53CM-coarse-graining-non-markovian-PhysRevA.95.032117, 54CM-quantum-critical-probing-PhysRevA.96.062117, 55CM-divisibility-of-quantum-dynamical-maps-PhysRevA.96.032111, 56CM-composite-CM-PhysRevA.96.032107, 57CM-time-delayed-quantum-feedback-PhysRevLett.115.060402, 58CM-OQD-delayed-coherent-feedback-Whalen_2017}. A closely related construction to what we follow here is the composite collision model \cite{56CM-composite-CM-PhysRevA.96.032107}, in which a subsystem of interest S and an intermediate environment E interact via an intra-system Hamiltonian, while E undergoes sequential collisions with independent ancillas representing a large external reservoir. The ancilla-driven dynamics is Markovian at the S$+$E level but can be non-Markovian at the level of S, and this framework has been used to derive non-Markovian master equations microscopically for S \cite{45Ciccarello-CM-non-Markovian-Incoherent-Swap_2013, 48CM-class-of-exact-memory-kernel-ME-PhysRevA.93.052111}.

The present work is structurally related to the composite collision model but departs from it in two essential respects. First, rather than partitioning the reservoir into independent ancillae acting on E, we model its net effect directly as an effective CPTP map on E between successive collisions. This allows us to access the full range of reservoir influence starting from complete memory preservation to full correlation erasure using a single unified framework. Second, rather than deriving a non-Markovian master equation for specific microscopic parameters, our objective is to understand the emergence of qualitatively distinct dynamical regimes like Markovian and non-Markovian, CP-divisible and CP-indivisible etc. as a direct consequence of the degree of correlation control exercised by the reservoir. In the following we construct and study a minimal model in which both S and E are single qubits with the reservoir acting on E via an effective CPTP maps between collisions. In the absence of the reservoir, a single qubit coupled to another qubit which acts the environment is an extremely non-Markovian system. 

We show that complete annihilation of the system-environment correlations requires a full reset of the environment and that this is the unique interaction-independent solution that ensures complete correlation-erasure. The strength of E to the reservoir governs a continuous transition from an identity map, which preserves correlations, to a complete reset, recovering the memory-less structure of a standard collision model. In the perfect-reset limit, the reduced dynamics exhibit a semi-group structure with a standard time-independent GKSL generator; resetting to an arbitrary state yields CP-divisible time-local dynamics \cite{17GKSL-Form_10.1063/1.522979, 4Davies1976-Open-quantum-book}. In the partial-reset regime, the competition between correlation generation and suppression produces both Markovian and non-Markovian dynamics depending on the model parameters.

Standard noise models, including generalized amplitude damping, and depolarizing processes, arise as natural physical realizations of this reservoir-induced partial-reset structure \cite{10NielsenChuangQuantum, 19All-open-dynamics,1Breuer-book}. As a concrete illustration, we analyze the model under a resonant excitation-exchange interaction with a reservoir-induced generalized amplitude damping channel, and demonstrate how the degree of environmental relaxation controls the Markovian or non-Markovian character of the reduced dynamics through direct manipulation of system--environment correlations.

The remainder of the paper is organised as follows. Section~\ref{sec:preliminaries} reviews the standard collision model, divisibility of dynamical maps, and the BLP and Choi-matrix measures used to classify the dynamics. Section~\ref{sec:model} introduces the two-qubit model, proves the uniqueness of the reset map, and identifies the dynamical regimes arising from partial environmental reset. 
Section~\ref{sec:perfect reset} derives exact closed-form dynamical maps in the perfect-reset limit and confirms CP-divisibility throughout. Section~\ref{sec:partial reset} analyses the partial-reset
regime for both the generalised depolarising and generalised amplitude-damping channels, characterising the three dynamical regimes numerically. Section~\ref{sec:conclusion} presents our conclusions.

\section{Collision models and non-Markovianity}
\label{sec:preliminaries}

We very briefly review some of the key ideas that are important for the rest of the paper in this section.

\subsection{Standard Collision Model}
\label{sec:prelim:SCM}

\begin{figure}[!htb]
    \centering
    \includegraphics[width=0.9\linewidth]{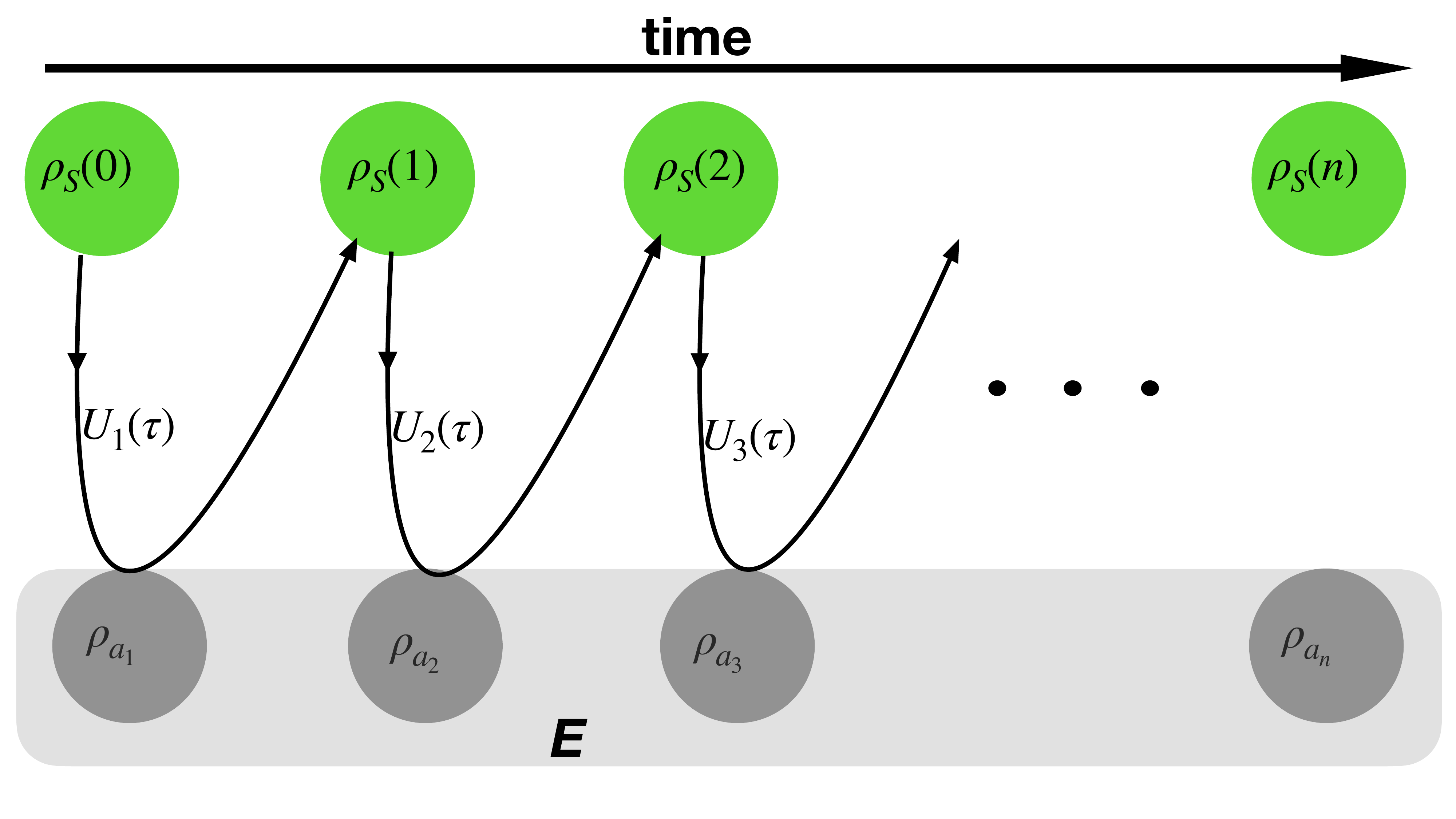}
    \caption{Schematic of the standard collision model. The environment E is partitioned into independent ancillae $a_i$, each interacting with S once via $U_{i}(\tau)$ before being discarded.}
    \label{fig:standard CM}
\end{figure}

 In the standard formulation of collision modes, which is schematically depicted in  Fig.~\ref{fig:standard CM}, the environment is partitioned into a continuous stream of independent ancillae, each initially prepared in state $\rho_{a_i}$ that interact or ``collide" sequentially with the system via a unitary $U_{i}(\tau)$. After each collision, the ancilla is either discarded or reset to its initial state and reused. The reduced density matrix of the system after the $n^{\rm th}$ collision is
\begin{equation}
    \rho_{\rm S} (n) = \mathrm{Tr}_{\rm E}\!\left[  U_{n}(\tau) \bigl(\rho_{\rm S}(n-1) \otimes \rho_{\rm E}\bigr) U_{n}^\dagger(\tau)\right],
\end{equation}
defining a single-step CPTP map $\Phi^\tau_i$ at each step. The overall $n$-step evolution is thus a composition of such maps,
\begin{equation}
    \mathcal{E}_{\rm S}(n) = \Phi^\tau_n\circ\cdots\circ\Phi^\tau_1,
\end{equation}
which, in the limit $\tau \to 0$, converges to a time-dependent GKSL dynamics \cite{17GKSL-Form_10.1063/1.522979,41CICCARELLO20221}  ( we use ``$\circ$" to denote the composition of two maps). Markovianity in this setting follows directly from the independence of the ancillae since each ancilla is freshly prepared and discarded after use, no memory is retained between steps. Note that in this case the state of each one of the ancillary systems that together make up the environment is the same and is denoted as $\rho_{\rm E}$.

The most important feature of collision models that we are interested in is that by construction, a microscopic picture is available for arriving at the dynamical map corresponding to each time step. The microscopic picture may not be realistic and in real experimental scenarios, the environment may be a composite system that does not break up into a collection of ancillary systems that interact sequentially with the system of interest. We still persist with collision models because they have been shown to produce master equations that model realistic systems quite accurately and more importantly, the microscopic picture can be constrained in such a way that a dynamical map always exists that describes the effect of each collision on the system.  

Non-Markovian effects are typically introduced by relaxing the assumption of independent ancillae, for instance, through ancilla--ancilla interactions, repeated use of the same ancilla, or pre-correlated environmental states \cite{44Ciccarello-CM-to-non-Markovian-PhysRevA.87.040103, 45Ciccarello-CM-non-Markovian-Incoherent-Swap_2013, 46Non-markovianity-system-env-correlation-CM-PhysRevA.89.052120, 47CM-All-optical-non-markovian-PhysRevA.91.012122, 48CM-class-of-exact-memory-kernel-ME-PhysRevA.93.052111, 49CM-for-non-Markovian-dynamics-PhysRevA.94.012106, 50CM-Non-Markovianity-and-coherence-and-system-environment-correlations-in-a-long-range-PhysRevA.96.022109, 51Cm-Rybar-indivisibleCM, 52CM-environmental-correlation-PhysRevA.90.032111, 53CM-coarse-graining-non-markovian-PhysRevA.95.032117, 54CM-quantum-critical-probing-PhysRevA.96.062117, 55CM-divisibility-of-quantum-dynamical-maps-PhysRevA.96.032111, 56CM-composite-CM-PhysRevA.96.032107, 57CM-time-delayed-quantum-feedback-PhysRevLett.115.060402, 58CM-OQD-delayed-coherent-feedback-Whalen_2017}. The present work pursues the second route. As mentioned earlier, we consider the minimal setting in which the system repeatedly interacts with the \textit{same} environmental qubit, and examine how the resulting accumulation of system--environment correlations shapes the reduced dynamics. In the modified framework introduced in Sec.~\ref {sec:model}, where the same environment qubit is reused, and a reservoir-induced map acts between collisions, the single-step map is no longer identical at each step. The resulting $n$-step dynamical map is denoted $\mathcal{E}_{\rm S}(n)$ throughout, reserving $\Phi^\tau_{\star}$ for a single step.

\subsection{Measure of Non-Markovianity}
\label{sec:prelim:BLP}

Since we are interested in systematically introducing environmental back action and non-Markovianity, it is important to quantify the memory effects. For this, we use the trace-distance-based measure introduced by Breuer, Laine, and Piilo (BLP) \cite{22BLP-measure-PhysRevLett.103.210401}. The measure uses the fact that trace distance between two states $\rho_{\rm S}^1$ and $\rho_{\rm S}^2$, defined as  $D(\rho_{\rm S}^1,\rho_{\rm S}^2) = \frac{1}{2}\|\rho_{\rm S}^1-\rho_{\rm S}^2\|_1$ with $ \|A\|_1 = \mathrm{Tr}\!\left[\sqrt{A^\dagger A}\right]$ is always contractive under memory-less (CPTP) evolution. Any transient increase in trace distance therefore signals a backflow of information from the environment to the system, and constitutes a signature of non-Markovian dynamics.

For a fixed pair of initial states $\rho_{\rm S}^{1,2}(0)$, the BLP information backflow is quantified as
\begin{equation}
    \mathcal{N}_{12}(\mathcal{E}_{\rm S}) = \sum_n  \max\!\left\{0,\;D_{n+1}-D_n\right\}, \label{sec:prelim:BLP_Non-markovianity for pair}
\end{equation}
where $D_n = D\!\left(\rho_{\rm S}^1(n),\rho_{\rm S}^2(n)\right)$. A measure of non-Markovianity is then obtained by taking the maximum over all initial state pairs,
\begin{equation}
    \mathcal{N}_{\mathrm{BLP}} =  \max_{\rho_{\rm S}^{1,2}(0)}
    \mathcal{N}_{12}(\mathcal{E}_{\rm S}). \label{sec:prelim:BLP_measure Maximised}
\end{equation}
For a the state of a qubit written in terms of its Bloch vector, 
\begin{equation}
    \rho_{\rm S} = \frac{1}{2}\!\left(\mathbb{1}_{\rm S} + \vec{a}\cdot\vec{\sigma}^S\right) \, \text{with} \; a_j=\mathrm{Tr[\sigma^S_j.\rho_{\rm S}]}, \label{sec:prelim:Bloch vector}
\end{equation}
where, $\sigma^S_j$'s are Pauli spin operators, the trace distance simplifies to
\begin{equation}
    D_n = \frac{1}{2}\left|\vec{a}_1^{(n)} -\vec{a}_2^{(n) } \right|,
\end{equation}
which means that the BLP measure can be evaluated directly from the Bloch-vector dynamics we obtain in Sec.~\ref{sec:partial reset}.

\subsection{Divisibility of dynamical maps}
\label{sec:prelim:maps}

Given a CPTP dynamical map $\mathcal{E}(t;0)$ that describe the finite-time open dynamics of a quantum system from time 0 to $t$, the map is divisible if we can write it as $\mathcal{E}(t;0)= \mathcal{V}(t;\tau) \circ \mathcal{E}(\tau;0)$ The dynamics is CP-divisible if, for all $\tau \in (0,t)$, the intermediate map, $\mathcal{V}(t;\tau)$ is also CPTP. The requirement of CP divisibility represents the strongest notion of Markovianity at the level of dynamical maps. A weaker condition, P-divisibility, requires only that $\mathcal{V}(t;\tau)$ be positive (positivity preserving, rather than completely positive) on the state space of the system. When the intermediate maps are not CPTP, but divisibility still holds, the dynamics is termed CP-indivisible~\cite{20Rivas_2014-Quantum_non-Markovianity-characterization, 29M.M.wolf-dividing_q-channel, 30Chruscinski-Markovian-criteria_2012, 31Chruscinski-divisibilty-Vs-backflow-PhysRevA.83.052128, 32Chruskinsky_divisibility, 33Chruskinsky-divisibilty-information-flow-non-invertible-maps-PhysRevLett.121.080407}.

In the continuous-time limit, CP-divisible dynamics with a time-independent generator reduces to a quantum dynamical semi group and one can construct an equivalent GKSL master equation with time-independent positive rates \cite{17GKSL-Form_10.1063/1.522979,19All-open-dynamics,4Davies1976-Open-quantum-book}. More generally, time-dependent rates yield CP-divisible dynamics with a time-local generator, while negative rates signal CP-indivisibility \cite{1Breuer-book,2Rivas-OQS-review,19All-open-dynamics}.

\subsection{Divisibility Measures}
\label{sec:prelim:div}

While the BLP measure effectively detects non-Markovianity via information backflow, it may fail to capture memory effects in weakly non-Markovian regimes where such backflow is not significant~\cite{21RHP-meaure-PhysRevLett.105.050403,34EternalNonMarkovian}. To obtain a finer classification, we complement the BLP measure with a divisibility-based diagnostic. Specifically, we use the BLP measure to distinguish P-indivisible from P-divisible dynamics, and employ the Choi-matrix representation of the intermediate map to further distinguish P-divisible from CP-divisible dynamics.

The intermediate map can be written explicitly as $\mathcal{V}(t;\tau) = \mathcal{E}(t;0) \circ \mathcal{E}^{-1}(\tau;0)$, provided the inverse exists. For a qubit, the associated Choi matrix (see Appendix.~\ref{appendix:choi}) is
\begin{equation}
    \mathcal{C}(t;\tau) = \left(\mathcal{I}_n \otimes \mathcal{V}(t;\tau)\right) \! \left[ \ket{\beta_{00}} \! \bra{\beta_{00}}\right],   \label{sec:prelim:Choi-matrix m2n}
\end{equation}
where $\ket{\beta_{00}} = (\ket{00}+\ket{11})/\sqrt{2}$ is a maximally entangled two qubit Bell state. Following the work of Rivas-Huelga-Plenio\cite{21RHP-meaure-PhysRevLett.105.050403}, the eigenvalues of $\mathcal{C}(t,\tau)$ directly encode the divisibility properties of the dynamics. To quantify deviations from CP-divisibility, we first discretize the time interval from 0 to $t$ into time steps of fixed duration $\delta t$ and labeled by $n$ with $t=n\delta t$ and define the measure
\begin{equation}
    \mathcal{N}_{\mathrm{div}} = \max_{n}\,\mathcal{N}_{\mathrm{div}}(n), \label{sec:prelim:Divisibility_measure}
\end{equation}
where
\begin{equation}
    \mathcal{N}_{\mathrm{div}}(n) = \sum_{i=1}^{4}\left|\mathrm{eig}_i\!\left(\mathcal{C}(n+1;n)\right)\right| - 1, \label{eq:Negative Choi}
\end{equation}
with $\mathrm{eig}_i\!\left(\mathcal{C}(n+1;n)\right)$ denoting the eigenvalues of the Choi matrix of the intermediate map from step $n$ to step $n+1$. A nonzero value of $\mathcal{N}_{\mathrm{div}}(n)$ signals the presence of a negative eigenvalue, indicating a failure of complete positivity at that step. Together, $\mathcal{N}_{\mathrm{div}}$ and the BLP measure provide a complete three-way classification of the dynamics, summarized in Table~\ref{tab:divisibility}.

\begin{table}[h]
\centering
\renewcommand{\arraystretch}{1.3}
\begin{tabular}{ccc}
\hline\hline
$\mathcal{N}_{\mathrm{div}}$ & $\mathcal{N}_{\mathrm{BLP}}$ & Classification \\
\hline
$= 0$ & $= 0$ & CP-divisible (Markovian) \\
$> 0$ & $= 0$ & CP-indivisible, P-divisible \\
$> 0$ & $> 0$ & CP-indivisible, non-P-divisible \\
\hline\hline
\end{tabular}
\caption{Classification of dynamics based on the divisibility measure 
$\mathcal{N}_{\mathrm{div}}$ and the BLP measure $\mathcal{N}_{\mathrm{BLP}}$.}
\label{tab:divisibility}
\end{table}

\section{Two qubit Model}
\label{sec:model}

The standard collision model leads to Markovian dynamics because the environment is reset or replaced after each collision, and correlations built up during one collision are discarded along with the ancilla before the next. Here we abandon this assumption. The system S repeatedly interacts with the same environment qubit E, so that correlations generated in one collision persist and influence all subsequent ones. As remarked earlier, in general, this would lead to a situation in which the dynamical map for the next time step may not even be defined. We therefore modulate the behaviour of the qubit E by coupling it to a larger system or reservoir that applies a dynamical map on qubit E after each collision. The main aim of this section is to identify what the reservoir-induced map acting on E must do to restore a controlled, closed description of the reduced dynamics, and to classify the dynamical regimes that arise when this requirement is met exactly or even partially.

\subsection{Joint Dynamics and Correlation Structure}
\label{sec:model:SE}  

We begin with the simplest setting: when there is no reservoir action, $\rm S$ and $\rm E$ interact unitarily in succession. We assume that they form a closed system — no external agent acts on either subsystem over the interaction interval $\tau$ during each collision —so the joint evolution is necessarily unitary, generated by the total Hamiltonian
\begin{equation}
    H_{\rm SE} = H_0 + H_I,
\end{equation}
where $H_0$ is the free part and $H_I$ the interaction. The corresponding unitary operator and superoperator are 
\begin{equation}
    U_{\rm SE}(\tau) = e^{-iH_{\rm SE}\tau}, \qquad
    \mathcal{U}_{\rm SE}(\tau)[\,\bullet\,]
    = U_{\rm SE}(\tau)\;\bullet\;U_{\rm SE}^\dagger(\tau),
\end{equation}
with $\hbar=1$. Were this unitary succession to continue indefinitely without any reservoir action, the joint state of S and E would remain well defined at every step. The difficulty is that a closed description for S alone breaks down after the first collision, as we now show. This breakdown is precisely what motivates the introduction of the reservoir map in Sec.~\ref {sec:model:reservoir}.

Starting from a product state $\rho_{\rm SE}(0) = \rho_{\rm S}(0) \otimes \rho_{\rm E}(0)$, the joint state after one collision is $\rho_{\rm SE}(1) = \mathcal{U}_{\rm SE}(\tau)[\rho_{\rm SE}(0)]$. It is useful to separate this state as follows into a product part and a correlation part,
\begin{equation}
    \rho_{\rm SE}(1) = \rho_{\rm S}(1) \otimes \rho_{\rm E}(1) 
    + \chi_{\rm SE}(1),
    \label{eq:decomposition}
\end{equation}
where $\rho_{\rm S(E)}(1) = \mathrm{Tr}_{\rm E(S)}[\rho_{\rm SE}(1)]$ are the reduced states and $\chi_{\rm SE}(1)=\rho_{\rm SE}(1)- \rho_{\rm S}(1) \otimes \rho_{\rm E}(1)$ encodes all correlations (both quantum and classical) generated by the interaction. In the standard collision model $\chi_{\rm SE}$ becomes irrelevant once the ancilla is discarded, but here it is not, and we modulate the correlation term by connecting a reservoir to qubit E.

After the first collision, since the joint state is no longer a product state a closed CPTP description of the reduced dynamics of S cannot always be constructed. The nonzero correlation term $\chi_{\rm SE}(1)$ means that the future evolution of S depends on the full joint state of S and E, not just $\rho_{\rm S}(1)$. Even if one were to remove the correlations, $\chi_{\rm SE}$, the state, $\rho_{\rm E}(1)$ depends on the initial state of S.  The next-step map for S depends on $\rho_{\rm S}(0)$ through $\rho_{\rm E}(1)$. This breaks linearity. To see this explicitly, consider two initial states $\rho_{\rm S}^1(0)$ and $\rho_{\rm S}^2(0)$ and a mixture of the two $\rho_{\rm S}^{12}(0) = p\rho_{\rm S}^1(0) + (1-p)\rho_{\rm S}^2(0)$.  After the first collision, linearity is preserved: $\rho_{\rm S}^{12}(1) = p\rho_{\rm S}^1(1) + (1-p)\rho_{\rm S}^2(1)$.  But after the second collision, the state of E entering that collision differs for each initial state, and so, in general,
\begin{equation}
    \rho_{\rm S}^{12}(2) \neq p\rho_{\rm S}^1(2) + (1-p)\rho_{\rm S}^2(2),
    \label{eq:nonlinear}
\end{equation}
unless the post-collision state of $\rm E$ is the same for all choices of the initial state of S or totally independent of it.  The main point of our work is to have minimal model wherein the correlations between $\rm S$ and $\rm E$ as well as the system dependence of the post-collision states of $\rm E$ can be systematically controlled and the conditions required for the existence of a viable description of the open dynamics of $\rm S$ in terms of dynamical maps corresponding to each collision be imposed in a systematic manner.

\subsection{Reservoir-Induced Map}
\label{sec:model:reservoir}

\begin{figure}[!htb]
    \centering
    \includegraphics[width=0.9\linewidth]{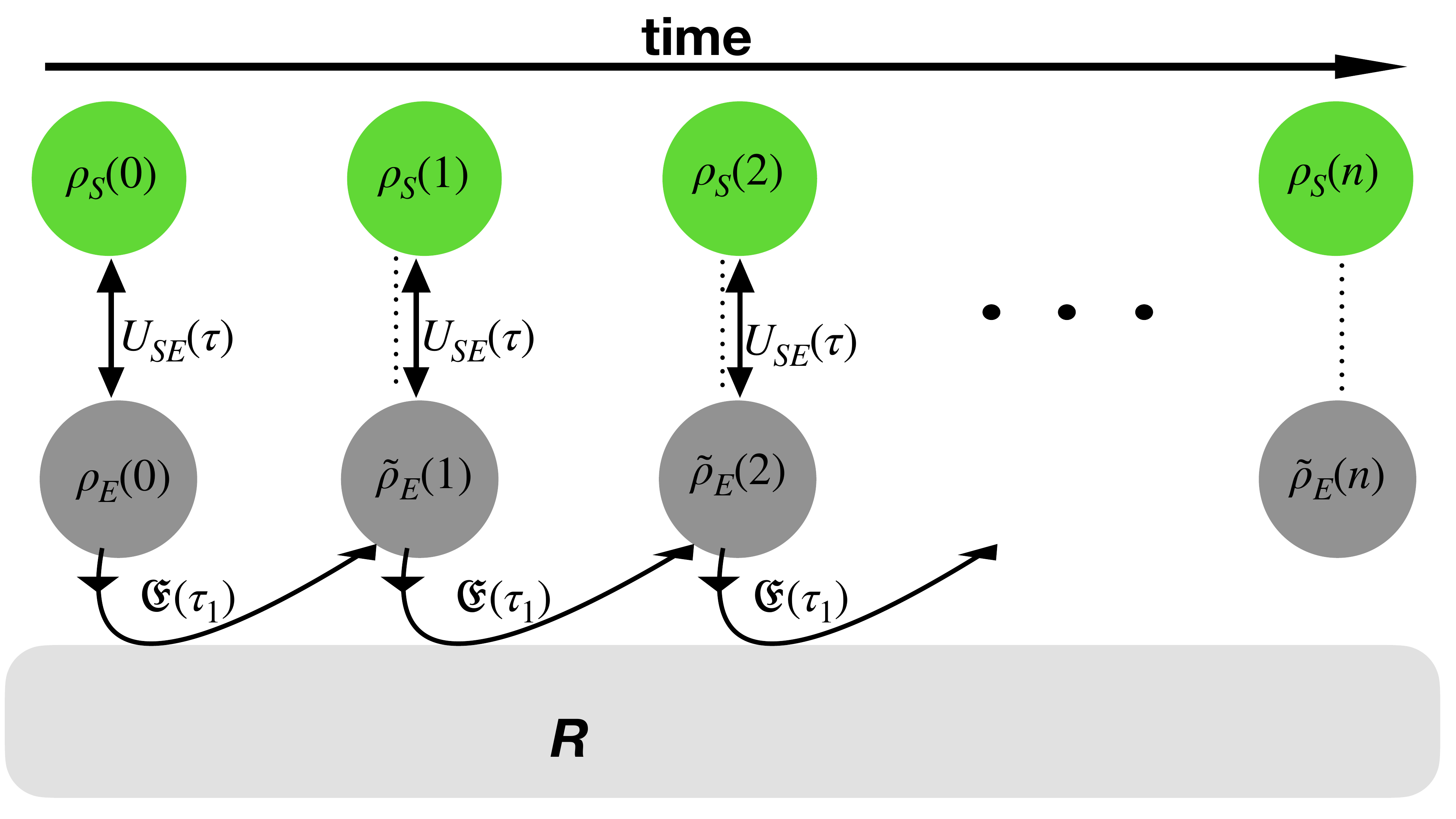}
    \caption{Schematic of the modified collision model. The system S repeatedly interacts with the same environment qubit E via $U_{\rm SE}(\tau)$, while a reservoir-induced CPTP map $\mathfrak{E}(\tau_1)$ acts on E between successive collisions. Unlike the standard construction, the environment is not discarded after each step, so system--environment correlations (dashed links) persist and accumulate, rendering the reduced dynamics of S history dependent.}
    \label{fig:Modified CM model}
\end{figure}

To control the state of E as well as to modulate $\chi_{\rm SE}$, we introduce a reservoir R that acts locally on E between successive S--E collisions via a CPTP map $\mathfrak{E}(\tau_1)$ over an interval $\tau_1$, as depicted in the Fig.~\ref{fig:Modified CM model}. We assume that while the map is being applied, both S and E do not undergo any other evolution. Alternatively, we can assume that the duration $\tau_1$ goes to zero. We define a single collision step as the composite of the two sequential operations: first, the unitary S--E interaction $\mathcal{U}_{\rm SE}(\tau)$, followed immediately by the reservoir-induced map $\mathfrak{E}(\tau_1)$ acting locally on E. Density matrices $\rho_{\rm SE}(n)$ denote the joint state right after the S--E interaction but \textit{before} the action of the reservoir, while $\tilde{\rho}_{\rm SE}(n)$ denote the joint state \textit{after} the reservoir has acted, which serves as the input to the next S--E collision. One complete time step in the evolution of S and E is given by, 
\begin{equation}
    \tilde{\rho}_{\rm SE}(n-1) \xrightarrow {\mathcal{U}_{\rm SE}(\tau)} \rho_{\rm SE}(n) \xrightarrow{\mathcal{I}_{\rm S} \otimes \mathfrak{E}(\tau_1)}  
    \tilde{\rho}_{\rm SE}(n),
    \label{eq:model one step}
\end{equation}
so that explicitly,
\begin{align}
    \rho_{\rm SE}(n)  & = \mathcal{U}_{\rm SE}(\tau) [\tilde{\rho}_{\rm SE}(n-1)],\nonumber \\
    \tilde{\rho}_{\rm SE}(n)  & = \mathcal{I}_{\rm S} \otimes \mathfrak{E}(\tau_1)  [\rho_{\rm SE}(n)].
    \label{sec:model:after one collision S-E_E-R}
\end{align}
Using the product-correlation decomposition Eq.~(\ref{eq:decomposition}), together with the definitions
\begin{equation}
    \tilde{\rho}_{\rm E}(n)  = \mathfrak{E}(\tau_1)[\rho_{\rm E}(n)], \; \tilde{\chi}_{\rm SE}(n)  = \mathcal{I}_{\rm S} \otimes \mathfrak{E}(\tau_1) [\chi_{\rm SE}(n)],
    \label{eq:tilded defs}
\end{equation}
where $\rho_{\rm S(E)}(n) = \mathrm{Tr}_{\rm E(S)}[\rho_{\rm SE}(n)]$ and $\chi_{\rm SE}(n) = \rho_{\rm SE}(n) - \rho_{\rm S}(n)\otimes\rho_{\rm E}(n)$ are the marginals and correlations immediately after the S--E interaction. The state after the reservoir acts takes the compact form
\begin{equation}
    \tilde{\rho}_{\rm SE}(n) = \rho_{\rm S}(n) \otimes \tilde{\rho}_{\rm E} (n)  + \tilde{\chi}_{\rm SE} (n).
    \label{eq:map decomposed}
\end{equation}

The first term shows that $\mathfrak{E}$ refreshes the environmental marginal to $\tilde{\rho}_{\rm E}(n)$, while the second term shows that it acts simultaneously on the accumulated correlations. Iterating the full composite step over $n$ collisions yields
\begin{equation}
    \tilde{\rho}_{\rm SE}(n) = \Big\{(\mathcal{I}_{\rm S} \otimes \mathfrak{E}(\tau_1)) \circ \mathcal{U}_{\rm SE}(\tau)\Big\}^n
    [\rho_{\rm SE}(0)],
    \label{eq:general n iterated S+E map}
\end{equation}
which induces an effective $n$-step dynamical map on S,
\begin{equation}
    \rho_{\rm S}(n) = \mathcal{E}_{\rm S}(n)[\rho_{\rm S}(0)].
    \label{eq:system map}
\end{equation}
Here $\mathcal{E}_{\rm S}(n)$ is in general not a composition of single-step CPTP maps which is in contrast to the standard collision-model in Sec.~\ref{sec:prelim:SCM} and this reflects the history dependence introduced by the accumulated correlations $\chi_{\rm SE}(n)$. Whether memory effects persist in $\mathcal{E}_{\rm S}(n)$ depends entirely on how $\mathfrak{E}$ acts on $\chi_{\rm SE}(n)$. We look at various possible actions of $\mathfrak{E}$ that can lead to well defined dynamical maps.

\subsection{Correlation Erasure and the Reset Condition}
\label{sec:model:erasure}

As noted earlier, for a closed, collision-model-like description to hold at each step, the correlation term in Eq.~(\ref{eq:map decomposed}) must vanish after the reservoir acts,
\begin{equation}
    (\mathcal{I}_{\rm S} \otimes \mathfrak{E}) [\chi_{\rm SE}(n)] = 0.
    \label{eq:erasure}
\end{equation}
Expanding $\chi_{\rm SE}(n)$ in the Pauli basis, this condition requires $\mathfrak{E}$ to annihilate every traceless environmental operator,
\begin{equation}
    \mathfrak{E}[\sigma_j^E] = 0, \quad \forall\, j.
    \label{eq:traceless}
\end{equation}

This condition has a natural interpretation in the broader landscape of collision models designed to generate or suppress memory. In the framework of Ciccarello, Palma, and Giovannetti (CPG) \cite{44Ciccarello-CM-to-non-Markovian-PhysRevA.87.040103,45Ciccarello-CM-non-Markovian-Incoherent-Swap_2013}, non-Markovian dynamics is introduced by inserting incoherent partial-swap operations between successive ancillas; these inter-ancillary collisions relay information about the system's past history forward through the bath, endowing the environment with memory in a controlled, dynamical way. The swap probability $p_{\rm S}$ interpolates between a standard memoryless model($p_{\rm S}=0$) and the fully non-Markovian extreme($p_{\rm S}=1$), in which the dynamics reduces to the continuous coherent interaction of S with a single ancilla. 

The present framework pursues the complementary question: rather than asking how much memory to inject into an initially uncorrelated bath, it asks what the reservoir must do to an already-correlated environment to erase that memory. Our model and the CPG construction therefore share the same non-Markovian extreme of  a single environmental qubit but arrive at it from opposite directions: CPG by maximising inter-ancillary information transfer, and the present work by setting the reservoir action to the identity.

For generic interactions, the unitary $U_{\rm SE}(\tau)$ generates correlations $\Gamma_{ij} = c_{ij} - a_i b_j$ that are spread across all Pauli directions simultaneously. The condition Eq.~(\ref{eq:traceless}) therefore requires $\mathfrak{E}$ to annihilate the entire traceless subspace of $\mathbb{S}(\mathcal{H}_{\rm E})$, not merely a subset of it. We show in Appendix~\ref{appendix:uniqueness} that the unique CPTP map satisfying the condition for arbitrary interactions is the complete reset,
\begin{equation}
    \mathfrak{E}[\rho_{\rm E}] = \eta_{\rm E},
    \label{eq:reset}
\end{equation}
which drives $E$ to a fixed state $\eta_{\rm E}$ regardless of its input. This leads to a product input $\rho_{\rm S}(n) \otimes \eta_{\rm E}$ at each subsequent collision and a well defined CPTP map exists for each collision from which a semi group evolution described by a GKSL master equation can be obtained.

\subsection{Partial Reset and Dynamical Regimes}
\label{sec:model:generic}

In practice, the reservoir need not achieve a perfect reset after each unitary collision. The general action of $\mathfrak{E}(\tau_1)$ on the Bloch sphere is
\begin{align}
    \mathfrak{E}(\tau_1)[\sigma_k^E] 
    &= \sum_j \mathfrak{E}_{jk}(\gamma_j\tau_1)\,\sigma_j^E, \\
    \mathfrak{E}(\tau_1)[\mathbb{1}_E] 
    &= \mathbb{1}_E 
    + \sum_j \mathfrak{E}_{j0}(\gamma_j\tau_1)\,\sigma_j^E,
    \label{eq:bloch map general}
\end{align}
where $\gamma_j>0$ are independent relaxation rates for each Bloch direction. The functions $\mathfrak{E}_{jk}$ and $\mathfrak{E}_{j0}$ are constrained only by complete positivity and the limiting conditions
\begin{equation}
    \mathfrak{E}_{jk}(0) = \delta_{jk},\,
    \mathfrak{E}_{jk}(\infty) = 0,\,
    \mathfrak{E}_{j0}(0)= 0,\,
    \mathfrak{E}_{j0}(\infty) = \eta_{{\rm E}k},
    \label{eq:boundary conditions}
\end{equation}
where $\eta_{{\rm E}k}$ are the Bloch components of the reservoir steady state $\eta_{\rm E}$. These conditions ensure that $\mathfrak{E}(\tau_1)$ interpolates between the identity ($\gamma_j \tau_1 \to 0$) and a complete  reset to $\eta_{\rm E}$ ($\gamma_j \tau_1 \to \infty$). A natural and consistent choice is
\begin{equation}
    \mathfrak{E}_{jk} (\gamma_j\tau_1) = \delta_{jk} e^{ -\gamma_j \tau_1}, \quad \mathfrak{E}_{j0} (\gamma_k\tau_1) = \eta_{{\rm E}k} (1 - e^{-\gamma_k\tau_1}).
    \label{eq:bloch map exponential}
\end{equation}

Three dynamical regimes follow directly. For $\gamma_j\tau_1\to 0$, $\mathfrak{E}$ approaches the identity so that the qubit E retains full memory of previous collisions, correlations accumulate, and the dynamics is strongly non-Markovian. For $\gamma_j\tau_1 \to \infty$, the map reduces to a complete reset to $\eta_{\rm E}$, recovering Markovian semi group dynamics. At intermediate values, a partial reset occurs: correlations are incompletely suppressed, and the dynamics may be Markovian or non-Markovian depending on the competition between the interaction strength and $\{\gamma_j\}$. These regimes are analyzed in 
Secs.~\ref{sec:perfect reset} and~\ref{sec:partial reset}.


\subsection{Physical Realizations}
\label{sec:model:realisations}

The abstract framework of the previous section admits several physical  realizations. The CPTP maps satisfying the limiting conditions given in Eqs.~(\ref{eq:boundary conditions}, \ref{eq:bloch map exponential}) fall into two broad classes depending on whether the suppression of the $\rm S-E$ correlations is 
uniform or non-uniform across Bloch directions  (see Appendix~\ref{appendix:uniqueness}). We present a physically realizable  example from each class.

\paragraph{Generalized depolarizing map:} The isotropic choice $\mathfrak{E}_{jk}=\lambda\,\delta_{jk}$ (see \ref{eq:correlation under R}) corresponds to the generalized depolarizing map
\begin{equation}
    \mathfrak{E}_{\rm GD}(\tau_1)[\rho_{\rm E}]  = \lambda\,\rho_{\rm E} + (1-\lambda)\,\eta_{\rm E}, \quad \lambda\equiv\lambda(\tau_1),
    \label{eq:gen dep map}
\end{equation}
with an arbitrary fixed point $\eta_{\rm E}$ within the Bloch ball. Since  $\mathfrak{E}_{jk}\propto\delta_{jk}$, the factor $\lambda$ factors through any  unitary $U_{\rm SE}$, and a closed-form effective dynamical map can be derived for  arbitrary $U_{\rm SE}$. The standard depolarizing channel is recovered as the special  case $\eta_{\rm E}=\mathbb{1}_{\rm E}/2$ with $\lambda=e^{-\gamma\tau_1}$, which serves as the  concrete physical example within this family. In Sec.~\ref{sec:partial reset} we  derive the general dynamical map for Eq.~(\ref{eq:gen dep map}) and specialise to  the depolarizing channel for an illustration.

\paragraph{Generalized amplitude damping (GAD).} The GAD channel models thermalization at rate $\gamma$ toward a reservoir with 
steady-state polarization $G_0\in[-1,1]$:
\begin{align}
    \mathfrak{E}_{\mathrm{GAD}}(\tau_1)[\sigma^{\rm E}_{1,2}]
    &= e^{-\gamma\tau_1}\,\sigma^{\rm E}_{1,2}, \nonumber\\
    \mathfrak{E}_{\mathrm{GAD}}(\tau_1)[\sigma^{\rm E}_3]
    &= e^{-2\gamma\tau_1}\,\sigma^{\rm E}_3, \nonumber\\
    \mathfrak{E}_{\mathrm{GAD}}(\tau_1)[\mathbb{1}_{\rm E}]
    &= \mathbb{1}_{\rm E}
       - G_0\!\left(1-e^{-2\gamma\tau_1}\right)\sigma^{\rm E}_3,
    \label{eq:physical GAD}
\end{align}
with fixed point 
\begin{equation}
   \eta_{\rm E}^{\mathrm{GAD}} = \frac{1}{2}(\mathbb{1}_{\rm E} - G_0 \sigma_3^{\rm E}), \label{eq:GAD fixed point} 
\end{equation}
recovering $|0\rangle\langle 0|$, $|1\rangle\langle 1|$, or $\mathbb{1}_E/2$ for 
$G_0=1,-1,0$ respectively. The contraction matrix  $\mathfrak{E}_{jk}^{\mathrm{GAD}} = \mathrm{diag}(e^{-\gamma\tau_1}, \, e^{-\gamma\tau_1}, \, e^{-2\gamma\tau_1})$ is anisotropic and does not factor through $U_{\rm SE}$ universally and so no closed-form dynamical map exists for arbitrary $U_{\rm SE}$. In  Sec.~\ref{sec:partial reset} we therefore fix a specific $U_{\rm SE}$ and derive the  resulting map directly, using the GAD channel to illustrate how anisotropic suppression  modifies the non-Markovian features relative to the generalized depolarizing case.

\section{Perfect Reset: Semi group and CP-Divisible Dynamics}
\label{sec:perfect reset}

As established in Sec.~\ref {sec:model:erasure}, complete correlation erasure at each step requires the reservoir to fully reset the environment to a fixed state $\eta_{\rm E}$ after every collision. In this section, we analyze the reduced dynamics that emerges in this idealized limit and show that it naturally gives rise to Markovian evolution, with the character of the dynamics whether it is semi group or merely  CP-divisible is determined by the reset state if it is fixed or varies from step to step.

\subsection{Time-Independent Reset and Semi group Dynamics}
\label{sec:perfect reset:TI}

When the environment is reset to the same state $\eta_{\rm E}$ after every collision, the joint state entering the $n^{\rm th}$ S-E interaction is
$\tilde{\rho}_{\rm SE}(n)=\rho_{\rm S}(n)\otimes\eta_{\rm E}$. With the initial
condition $\rho_{\rm SE}(0)=\rho_{\rm S}(0)\otimes\rho_{\rm E}(0)$, the single-step reduced map
\begin{equation}
    \Phi_{\star}^\tau[\rho_{\rm S} ]  = \mathrm{Tr}_{\rm E} \Big[ U_{\rm SE}(\tau) (\rho_{\rm S} \otimes\star)\, U_{\rm SE}^\dagger(\tau)
    \Big]
    \label{eq:single step map def TI}
\end{equation}
leads to the $n$-step dynamics
\begin{equation}
    \mathcal{E}_{\rm S}(n) = \bigl(\Phi_{\eta_{\rm E} }^\tau\bigr)^{n-1}
      \circ\,\Phi_{\rho_{\rm E}(0)}^\tau.
    \label{eq:discrete semigroup}
\end{equation}
Since each factor is CPTP by construction, the intermediate map $\mathcal{V}_{\rm S} (n,m) = (\Phi_{\eta_{\rm E} }^\tau)^{n-m}$ is CPTP for all
$n \geq m \geq 0$, so~\eqref{eq:discrete semigroup} is CP-divisible throughout. When $\rho_{\rm E}(0)=\eta_{\rm E}$ the first factor coincides with the rest and the dynamics reduces to a strict discrete semi group $\mathcal{E}_{\rm S}(n)=(\Phi_{\eta_{\rm E}}^\tau)^n$ recovering the standard case with the Born and Markov approximations in place.

\subsection{Example: Resonant Excitation-Exchange Interaction
            (Time-Independent Reset)}
\label{sec:perfect reset:example TI}

For S and E we have the Hamiltonian, $H_{\rm SE} = H_0 + H_I$, where 
\begin{align}
    H_0 &= \frac{\omega_0}{2}  \big(\sigma_3^S \otimes \mathbb{1}_{\rm E}
    + \mathbb{1}_{\rm S} \otimes \sigma_3^{\rm E} \big), \nonumber \\
    H_I &= \alpha\big(\sigma_1^S \otimes \sigma_1^E + \sigma_2^{\rm S} \otimes \sigma_2^{\rm E}\big) \nonumber\\
   & = 2\alpha\big(\sigma_+^{\rm S} \otimes \sigma_-^{\rm E}  + \sigma_-^{\rm S} \otimes \sigma_+^{\rm E} \big). \label{eq:XX-YY_total}
\end{align}
Since $[H_0, H_I] = 0$, the free evolution commutes with the interaction, and the dynamics can be analyzed entirely in the interaction picture without loss of generality. The initial state of E is,
\begin{equation}
    \rho_{\rm E}(0)=\tfrac{1}{2} \left(\mathbb{1}_{\rm E} + b^{(0)}_j \sigma^{\rm E}_j \right),
    \label{eq:initial env state}
\end{equation}
which resets after each collision to the fixed longitudinal state
\begin{equation}
    \eta_{\rm E}=\tfrac{1}{2} \left(\mathbb{1}_{\rm E} - G_0\,\sigma_3^{\rm E} \right),
    \quad G_0\in[-1,1].
    \label{eq:reset_state_TI}
\end{equation}
The single-step map for a general environmental state is derived in Appendix~\ref{appendix:single-step map}. Substituting the initial state~\eqref{eq:initial env state} and the reset
state~\eqref{eq:reset_state_TI} into Eq.~\eqref{eq:app:single-step map}
gives $\Phi^\tau_{\rho_{\rm E}(0)}$ and $\Phi^\tau_{\eta_{\rm E}}$ respectively.
Evaluating the composition~\eqref{eq:discrete semigroup} as shown in
Appendix~\ref{appendix:perfect reset derivation fixed pt}, the
$n$-step dynamical map in the ordered basis $\{\mathbb{1}_{\rm S} , \sigma^{\rm S} _1, \sigma^{\rm S}_2, \sigma^{\rm S}_3\}$ is
\begin{equation}
    \mathcal{E}_{\rm S}(n) \!  = \! 
      {\small{ \begin{pmatrix} 
        1 & 0 & 0 & 0 \\
        0 & c^n & 0 & b^{(0)}_2 \! s  c^{n-1} \\
        0 & 0 & c^n & -b^{(0)}_1 \! s c^{n-1} \\
        f_{\mathrm{TI}}(n) & -b^{(0)}_2 \! s c^{2n-1}
        & b^{(0)}_1 \! s c^{2n-1} & c^{2n} 
    \end{pmatrix}}}  ,
    \label{eq:dynamical map TI}
\end{equation}
where $c \equiv \cos(2\alpha\tau)$, $s \equiv \sin(2\alpha\tau)$, and
\begin{equation}
    f_{\mathrm{TI}}(n) = b^{(0)}_3 s^2 c^{2(n-1)} - G_0\!\left(1-c^{2(n-1)}\right).
    \label{eq:fTI}
\end{equation}
In Bloch-vector notation, the map acts as
\begin{equation}
    \vec{a}^{(n)}
    =
    \begin{pmatrix}
        c^n\,a_1^{(0)} + b^{(0)}_2\,s\,c^{n-1}\,a_3^{(0)} \\[4pt]
        c^n\,a_2^{(0)} - b^{(0)}_1\,s\,c^{n-1}\,a_3^{(0)} \\[4pt]
        c^{2n}\,a_3^{(0)} + f_{\mathrm{TI}}(n)
    \end{pmatrix}.
    \label{eq:bloch TI}
\end{equation}
The transverse components carry a contribution from the initial longitudinal polarization $a_3^{(0)}$ whenever $b^{(0)}_1$ or $b^{(0)}_2$ is non-zero, reflecting the mixing induced by the off-diagonal elements of $\Phi^\tau_{\rho_{\rm E}(0)}$ at the first step. From the second step onward, the purely longitudinal reset state decouples the two sectors, so those off-diagonal entries decay as $c^{n-1}$ and $c^{2n-1}$ and vanish as $n\to\infty$. The longitudinal component relaxes toward the fixed point $-G_0$ set by the reset state, with the steady state $\vec{a}^{(\infty)}=(0,0,-G_0)$ independent of
the initial system state.

When the initial environment state is also longitudinal, $b^{(0)}_1 = b^{(0)}_2 = 0$, the map~\eqref{eq:dynamical map TI} is given by a diagonal matrix, 
\begin{equation}
    \mathcal{E}_{\rm S}(n)\big|_{b^{(0)}_{1,2}=0}  =
    \begin{pmatrix}
        1 & 0 & 0 & 0 \\
        0 & c^n & 0 & 0 \\
        0 & 0 & c^n & 0 \\
        f_{\mathrm{TI}}(n) & 0 & 0 & c^{2n}
    \end{pmatrix}.
    \label{eq:dynamical map TI diag}
\end{equation}
Note that if $\rho_{\rm E}(0 ) =\eta_{\rm E}$, i.e.~$b^{(0)}_3 = -G_0$, then $f_{\mathrm{TI}}(n)=-G_0(1-c^{2n})$ and the map is a pure semi group as expected. The diagonal structure of~\eqref{eq:dynamical map TI diag} makes CP-divisibility manifest and it is easy to see that its inverse exists. The intermediate map $\mathcal{V}_{\rm S} (n,m) = \mathcal{E}_{\rm S}(n-m)$ is CPTP for all $n\geq m\geq 0$.

\subsection{Time-Dependent Reset and CP-Divisible Dynamics}
\label{sec:perfect reset:TD}

We now allow the reset state to vary from step to step,
$\eta_{\rm E}\to\eta_{\rm E}^{(n)}$. Each collision induces a distinct single-step
map corresponding to a distinct environmental state~\eqref{eq:single step map def TI} and the $n$-step dynamics is the ordered composition
\begin{equation}
\mathcal{E}_S(n)
= \Phi^\tau_{\eta_{\rm E}^{(n-1)}} \circ \cdots \circ \Phi^\tau_{\eta_{\rm E}^{(1)}} \circ \Phi^\tau_{\rho_{\rm E}(0)}.
\label{eq:composition TD}
\end{equation}
where the first factor $\Phi^\tau_{\rho_{\rm E}(0)}$ uses the general initial environment state~\eqref{eq:initial env state} and all subsequent factors use the varying reset state~\eqref{eq:reset_state_TD}. Since every factor is CPTP, the intermediate map, 
\[ \mathcal{V}_{\rm S}(n;m)=\Phi^\tau_{\eta_{\rm E}^{(n-1)}} \circ \cdots \circ \Phi^\tau_{\eta_{\rm E}^{(m+1)}}, \] 
is CPTP for all $n>m\geq 0$, so the dynamics is CP-divisible throughout.

\subsection{Example: Time-Dependent Reset for the Resonant Interaction}
\label{sec:perfect reset:example TD}

For the Hamiltonian in Eq.~\eqref{eq:XX-YY_total}, we take the reset state at step
$n$ to be a longitudinal qubit state with a step-dependent polarization,
\begin{equation}
    \eta_{\rm E}^{(n)}   = \tfrac{1}{2}\!\left(\mathbb{1}_E+G_0^{(n)}\,\sigma_3^{\rm E} \right), \quad G_0^{(n)}\in[-1,1],
    \label{eq:reset_state_TD}
\end{equation}
where $G_0^{(n)} \equiv G_0(n\tau)$ is a prescribed discrete sequence. This simulates a situation where the joint evolution of S and E changes the states of both qubits at each collision, and the reservoir only removes the correlations between the two without resetting the state of E. Evaluating the composition~\eqref{eq:composition TD} as shown in Appendix~\ref{appendix:perfect reset derivation varying pt}, the $n$-step dynamical map is
\begin{equation}
    \mathcal{E}_{\rm S}(n)  =
    {\footnotesize{ \begin{pmatrix}
        1 & 0 & 0 & 0 \\
        0 & c^n & 0 & b^{(0)}_2\! s c^{n-1} \\
        0 & 0 & c^n & -b^{(0)}_1\! s c^{n-1} \\
        f_{\mathrm{TD}}(n) & -b^{(0)}_2 \! s c^{2n-1}
        & b^{(0)}_1 \! s c^{2n-1} & c^{2n}
    \end{pmatrix}}},
    \label{eq:dynamical map TD}
\end{equation}
where
\begin{equation}
    f_{\mathrm{TD}}(n) = b^{(0)}_3 s^2 c^{2(n-1)} + s^2\sum_{m=0}^{n-2} G_0^{(n-1-m)} c^{2m}.
    \label{eq:fTD}
\end{equation}

The structure of~\eqref{eq:dynamical map TD} mirrors that of the time-independent map~\eqref{eq:dynamical map TI} and the off-diagonal transverse entries arising from the general initial state persist with the same $c^{n-1}$ and $c^{2n-1}$ decay, while all time-dependence of the varying reset sequence $\{G_0^{(n)}\}$ enters exclusively through the scalar function $f_{\mathrm{TD}}(n)$. This is a direct consequence of the purely longitudinal reset state~\eqref{eq:reset_state_TD}: a reset state with non-zero transverse polarization would additionally couple the transverse sector at every step.

When $b^{(0)}_1 = b^{(0)}_2=0$ the map becomes diagonal and reduces to Eq.~\eqref{eq:dynamical map TD} of the main text. Setting $G_0^{(n)}=-G_0$ for all $n$ collapses $f_{\mathrm{TD}}$ to $f_{\mathrm{TI}}$, recovering the time-independent map~\eqref{eq:dynamical map TI} exactly.

The CP-divisibility of~\eqref{eq:dynamical map TD} is confirmed by computing the Choi matrix of the single-step intermediate map $\mathcal{V}_S (n+1; n) = \Phi^\tau_{\eta^{(n)}_{\rm E}}$ via the method of Sec.~\ref{sec:prelim:div} in its discrete form. The eigenvalues of the Choi matrix are
\begin{equation}
    \left\{
        \frac{s^2\!\left(1\pm G_0^{(n)}\right)}{2},\;
        \frac{1+c^2\pm\sqrt{s^2\!\left(G_0^{(n)}\right)^2+4c^2}}{2}
    \right\},
    \label{eq:choi eigs TD}
\end{equation}
which are all non-negative for $|G_0^{(n)}|\leq 1$, confirming that every single-step map is CP and hence the full dynamics is CP-divisible regardless of the choice of sequence $\{G_0^{(n)}\}$.

The perfect-reset scenario thus provides a clean reference class of Markovian dynamics parametrized entirely by the reset-state sequence $\{G_0^{(n)}\}$ and the interaction Hamiltonian. The first collision, governed by the general initial state $\rho_{\rm E}(0)$, imprints transient off-diagonal elements on the map; these decay geometrically and leave no trace on the long-time dynamics. All transient environmental effects are encoded in the scalar function $f_{\mathrm{TD}}(n)$, a discrete causal convolution of $\{G_0^{(n)}\}$ with the geometric kernel $c^{2m}$. We now move beyond this idealized scenario and consider the case where the reservoir induces only a partial reset, allowing
correlations to persist and generate the non-Markovian dynamics analyzed.

\section{Partial Reset and Non-Markovian Dynamics}
\label{sec:partial reset}

When the reservoir induces only a partial reset of the environment, the correlation term $(\mathcal{I}_{\rm S} \otimes \mathfrak{E} (\tau_1)) [\chi_{\rm SE}(n)]$ does not vanish between collisions. Correlations generated during the $n^{\rm th}$ S--E interaction persist into the $(n+1)^{\rm th}$ collision, making the reduced dynamics of S, history dependent. As established in Sec.~\ref {sec:model:generic}, the degree of correlation suppression is controlled by the parameters $\gamma_j \tau_1$. Weak reservoir action (small $\gamma_j\tau_1$) leaves correlations largely intact and produces strong memory effects, while strong reservoir action (large $\gamma_j\tau_1$) approaches the perfect reset limit of Sec.~\ref{sec:perfect reset}. The intermediate regime is governed by a competition between correlation generation through $U_{\rm SE}(\tau)$ and correlation suppression through $\mathfrak{E}(\tau_1)$, and we study the non-Markovian dynamics that arise in this regime using  a concrete model that admits full analytical treatment.

\subsection{Generalized Depolarizing Channel}
\label{sec:partial reset:uniform contraction}

The generalized depolarizing (GD) channel acts on the environment by uniformly contracting all correlations toward a fixed point $\eta_{\rm E}$ as discussed in Sec.~\ref{sec:model:realisations} and made explicit in Eq.~\eqref{eq:correlation under R}. Its action on the environmental state and the correlation matrix $\chi_{\rm SE}$ is
\begin{align}
    \rho_{\rm E} & \; \longrightarrow \;  \mathfrak{E}_{\mathrm{GD}}[\rho_{\rm E}] = \lambda\,\rho_{\rm E} + (1-\lambda)\,\eta_{\rm E},
    \label{eq:action of GD}\\
    \chi_{\rm SE} &\;\longrightarrow\; \left(\mathcal{I}_{\rm S} \otimes \mathfrak{E}_{\mathrm{GD}}\right)[\chi_{\rm SE}]
        = \lambda\,\chi_{\rm SE},
    \label{eq:action of GD on correlation}
\end{align}
where $\lambda\in[0,1]$ controls the strength of the reset with $\lambda=0$ recovering the perfect reset and $\lambda=1$ leaves the environment unchanged.

Starting from the initial product state $\rho_{\rm SE}(0) = \rho_{\rm S}(0) \otimes \rho_{\rm E} (0)$, the GD channel acts between
successive S--E collisions (Fig.~\ref{fig:Modified CM model}). Using the definitions~\eqref{eq:model one step}, the joint state after the first collision and the subsequent GD reset is
\begin{align}
    \rho_{\rm SE}(1)  &= \mathcal{U}_{\rm SE}\bigl[\rho_{\rm SE}(0)\bigr], \\
    \tilde{\rho}_{SE}(1) &= \bigl( \mathcal{I}_{\rm S}\otimes \mathfrak{E}_{\mathrm{GD}} \bigr) \bigl[ \rho_{\rm SE}(1) \bigr] \nonumber\\
    & = \lambda\,\rho_{\rm SE}(1) + (1-\lambda)\,\rho_{\rm S}(1)\otimes\eta_{\rm E}.
    \label{eq:rho tilde 1}
\end{align}
The second equality follows directly from Eqs.~\eqref{eq:action of GD}-\eqref{eq:action of GD on correlation} showing that the GD channel contracts the correlation part by $\lambda$ and replaces the environmental marginal by the convex combination $\lambda\rho_{\rm E}(1)+(1-\lambda)\eta_{\rm E}$, so the full joint state becomes a convex mixture of the correlated post-collision state and a product state with the fixed-point environment. Iterating this argument through $n$ steps yields
\begin{equation}
    \tilde{\rho}_{\rm SE}(n)
    = \lambda^n\,\rho_{\rm SE}(n)
      + (1-\lambda^n)\,\rho_{\rm S}(n)\otimes\eta_{\rm E},
    \label{eq:rho tilde n - uniform - GD}
\end{equation}
where $\rho_{\rm SE}(n)=(\mathcal{U}_{\rm SE})^n[\rho_{\rm SE}(0)]$ is the fully correlated state that would result from $n$ unitary collisions without any reset. Taking the partial trace over $E$, the reduced $n$-step dynamical map on the system is
\begin{equation}
    \mathcal{E}_{\rm S}(n)  = \lambda^n\,\mathcal{E}^{(1)}_{\rm S} (n) + (1-\lambda^n)\,\mathcal{E}^{(0)}_{\rm S}(n),
    \label{eq:GD convex map}
\end{equation}
where the two constituent maps are
\begin{align}
    \mathcal{E}^{(1)}_{\rm S}(n)[\rho_{\rm S}]  & = \mathrm{Tr}_{\rm E} \!\left[  (\mathcal{U}_{\rm SE})^n[\rho_{\rm SE}(0)]  \right], \label{eq:E1}\\
    \mathcal{E}^{(0)}_S(n)
    &= \bigl(\Phi^\tau_{\eta_{\rm E}}\bigr)^{n-1} \circ\,\Phi^\tau_{\rho_{\rm E}(0)}.
    \label{eq:E0}
\end{align}
Here we have suppressed the collision time $\tau$ for brevity, and used the fact that $(\mathcal{U}_{\rm SE}(\tau))^n=\mathcal{U}_{\rm SE}(n\tau)$.

The map $\mathcal{E}^{(0)}_S(n)$ is precisely the perfect-reset CP-divisible map derived in Appendix~\ref{appendix:perfect reset derivation fixed pt}, while $\mathcal{E}^{(1)}_{\rm S}(n)$ is the coherent unitary map that accumulates all S--E correlations without any reset. Since both constituents are CPTP, the full map~\eqref{eq:GD convex map} is also CPTP for all $n$ because it is a convex combination of CPTP maps with non-negative weights $\lambda^n$ and $1-\lambda^n$ that sum to unity. It may be noted that arbitrary sums or differences of CPTP maps however, need not be CPTP. 

The divisibility character of~\eqref{eq:GD convex map} is, however, non-trivial. The map $\mathcal{E}^{(0)}_{\rm S}(n)$ is CP-divisible by construction, but $\mathcal{E}^{(1)}_{\rm S}(n)$ generically is not, since repeated unitary collisions build up S--E correlations that cannot be undone by a subsequent CP map. The weight $\lambda^n$ governing the contribution of $\mathcal{E}^{(1)}_{\rm S} (n)$ decays geometrically with the number of collisions, so the degree of non-Markovian character is controlled directly by $\lambda$. For $\lambda$ close to zero the dynamics is dominated by the CP-divisible component and is nearly Markovian, while for $\lambda$ close to unity the coherent component dominates and strong memory effects emerge. We analyze the Markovian to non-Markovian and CP to non-CP divisibility regime in the next section numerically using the measures defined in Sec.~\ref{sec:prelim:BLP} and Sec.~\ref{sec:prelim:div}.

\subsection{Dynamical Regimes: Uniform Contraction}
\label{sec:partial reset:dyn regime-uniform}

We now evaluate the dynamical map~\eqref{eq:GD convex map} explicitly for the resonant XX-YY  interaction given in Eq.~\eqref{eq:XX-YY_total} and the initial environment state given in Eq.~\eqref{eq:initial env state} with $b_1^{(0)}=1$, $b_2^{(0)}=b_3^{(0)}=0$, and $G_0=1$. The map $\mathcal{E}^{(1)}_{\rm S}(n)$ describes $n$ unitary S--E collisions with no reservoir action. Since $\bigl(\mathcal{U}_{\rm SE} (\tau)\bigr)^n = \mathcal{U}_{\rm SE}(n\tau)$, it is obtained from the single-step map given in Eq.~\eqref{eq:app:single-step map} of Appendix~\ref{appendix:single-step map} by the substitution $\tau\to n\tau$, i.e.\ $c\to c_n\equiv\cos(2\alpha n\tau)$ and $s\to s_n\equiv\sin(2\alpha n\tau)$ as
\begin{equation}
    \mathcal{E}^{(1)}_{\rm S}(n) =
    \begin{pmatrix}
        1 & 0 & 0 & 0 \\
        0 & c_n & 0 & 0 \\
        0 & 0   & c_n & -s_n \\
        -s_n^2 & 0 & c_n s_n & c_n^2
    \end{pmatrix},
    \label{eq:E1 explicit}
\end{equation}
where we have substituted the values of $b_j^{(0)}$. The map $\mathcal{E}^{(0)}_{\rm S}(n)$ is the perfect-reset CP-divisible map derived in Appendix~\ref{appendix:perfect reset derivation fixed pt}. The full $n$-step dynamical map is the convex combination (in Eq.~(\ref{eq:GD convex map}))
\begin{equation}
    \mathcal{E}_{\rm S}(n) = \lambda^n\,\mathcal{E}^{(1)}_{\rm S}(n)  + (1-\lambda^n)\,\mathcal{E}^{(0)}_{\rm S}(n).
\end{equation} 
The boundaries $s=0$ and $\lambda=0$ are Markovian by construction since $s=0$ implies $U_{\rm SE} = \mathbb{1}_{\rm SE}$, while $\lambda=0$ recovers the CP-divisible semigroup of Sec.~\ref{sec:perfect reset}.

\begin{figure}[!htb]
    \centering
    \includegraphics[width=0.9\linewidth]{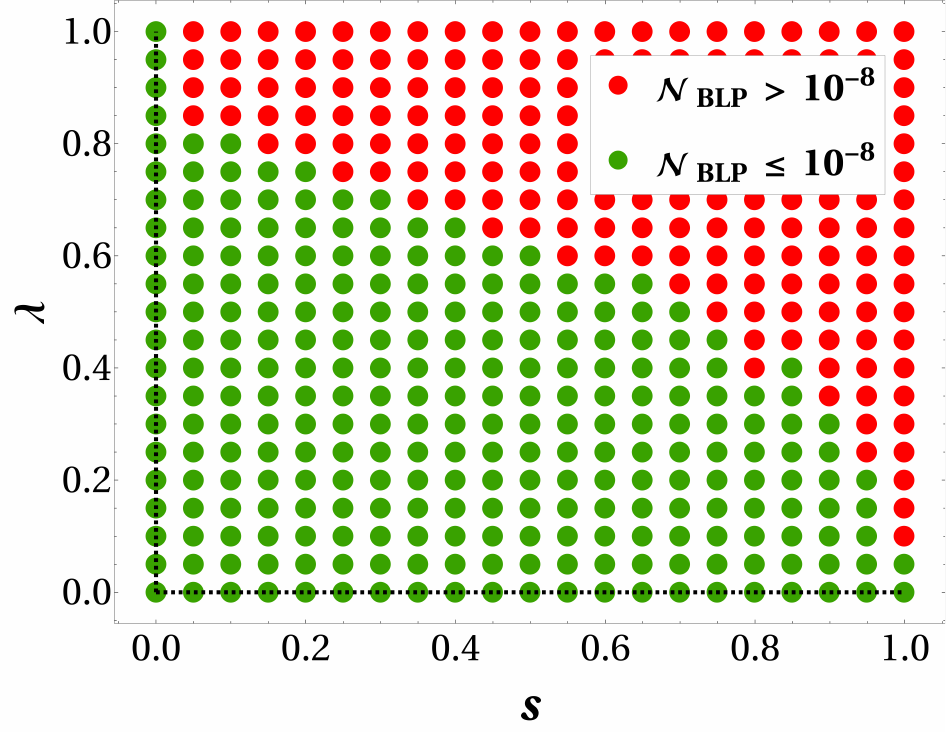}
    \caption{BLP non-Markovianity measure $\mathcal{N}_{\mathrm{BLP}}$ in the $(s,\lambda)$ parameter space, evaluated over $n=100$ collisions with $b_1^{(0)}=1$, $b_2^{(0)}=b_3^{(0)}=0$, and $G_0=1$. The green dots denotes points on $(s, \lambda)$ plane for which the dynamics is Markovian while the red dots denote the non-Markovian regime where $\mathcal{N}_{\mathrm{BLP}}$ is greater than a small threshold value of $10^{-8}$. The measure vanishes along the dashed boundaries $s=0$ and $\lambda=0$, which are Markovian by construction. Non-Markovian dynamics emerges above when $\lambda > \lambda^*(s)$ with $\lambda^*(s)$ traced out by the boundary between the green and red dots in the figure. We see that $\lambda^*(s)$  decreases with increasing interaction strength.}
    \label{fig:BLP-uniform}
\end{figure}

The parameter space we choose is $(s,\lambda)$ where $s=\sin(2\alpha n \tau)$ depends on the interaction strength in Eq.~\eqref{eq:XX-YY_total} and the number of collisions $n$. We keep $n$ fixed and consider the range $0\leq s \leq 1$ which corresponds to $0 \leq \alpha \leq \pi/(4 n \tau)$ considering the periodicity of the sine function. We can classify the open dynamics of S into three regimes in this parameter space according to the measures of Sec.~\ref{sec:preliminaries}, evaluated over $n=100$ collisions. Fig.~\ref{fig:BLP-uniform} shows the non-Markovianity measure, $\mathcal{N}_{\mathrm{BLP}}$ across the full parameter space. The measure vanishes along $s=0$ and $\lambda=0$ by construction, and remains zero in a neighborhood of these boundaries up to a threshold value that we choose as $10^{-8}$. Non-Markovian dynamics emerges when $ \lambda > \lambda^*(s)$, where $\lambda^*(s)$ is a function that decreases with increasing interaction strength, tracing a phase boundary in the $(s,\lambda)$ plane.

\begin{figure}[!htb]
    \centering
    \includegraphics[width=0.9\linewidth]{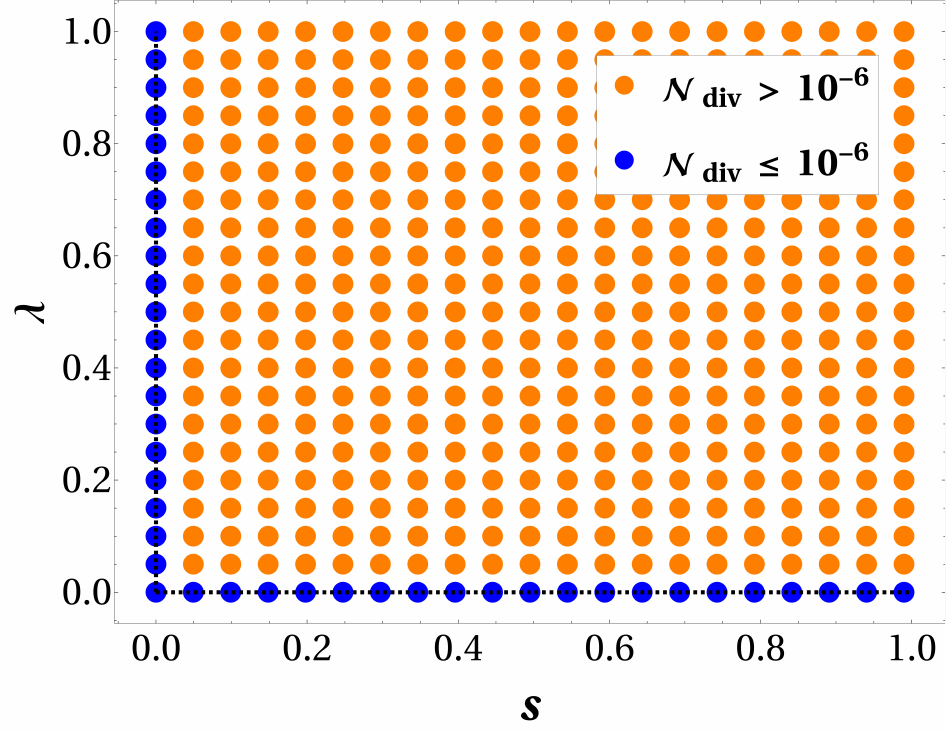}
    \caption{Divisibility measure $\mathcal{N}_{\mathrm{div}}$ in the $(s,\lambda)$ parameter space for the same parameters as Fig.~\ref{fig:BLP-uniform}. The witness is nonzero throughout the interior $s,\lambda>0$, including in regions where $\mathcal{N}_{\mathrm{BLP}}=0$, revealing a CP-indivisible yet P-divisible regime between the two phase boundaries.}
    \label{fig:Ndiv-uniform}
\end{figure}

\begin{figure}[!htb]
    \centering
    \includegraphics[width=0.9\linewidth]{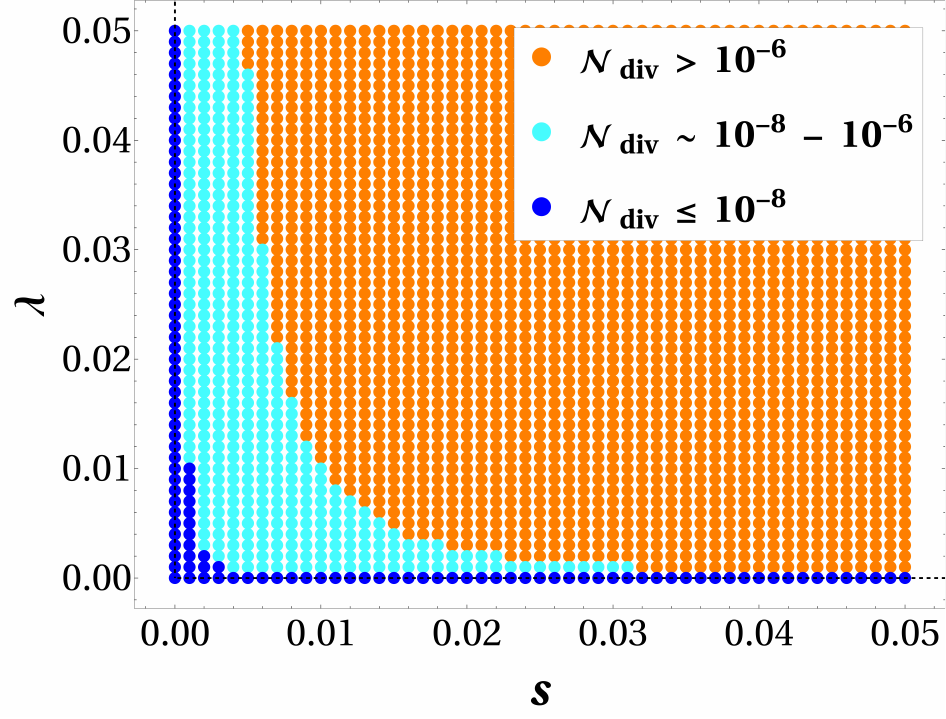}
    \caption{Divisibility measure $\mathcal{N}_{\mathrm{div}}$ in the restricted region $s,\lambda\in[0,0.05]$, revealing the onset of CP-indivisibility.}
    \label{fig:Ndiv-uniform-small}
\end{figure}

Fig.~\ref{fig:Ndiv-uniform} shows $\mathcal{N}_{\mathrm{div}}$ over the same parameter space. The witness is nonzero throughout the interior $s,\lambda>0$, including in regions where $\mathcal{N}_{\mathrm{BLP}}=0$. The region between the onset of $\mathcal{N}_{\mathrm{div}}$ and the BLP phase boundary therefore contains dynamics that are CP-indivisible yet P-divisible, which represent an intermediate class of dynamics (see Table~\ref{tab:divisibility}). To obtain a clearer picture of the onset of CP-indivisibility, we zoom into the restricted region  $s,\lambda \in [0,0.05]$, shown in Fig.~\ref{fig:Ndiv-uniform-small}. We see that the reservoir coupling can influence ${\mathcal N}_{\rm div}$ when $s$ is small, but it has no influence when the coupling strength increases. 

\begin{figure}[!htb]
    \centering
    \includegraphics[width=0.9\linewidth]{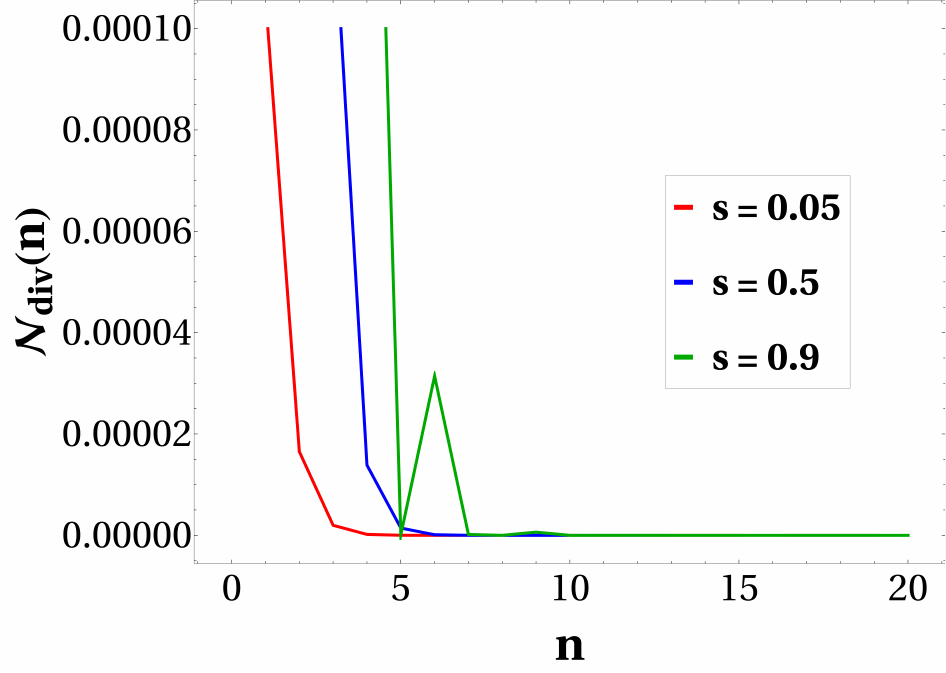}
    \caption{Step-wise CP witness $\mathcal{N}_{\mathrm{div}}(n)$ as a function of collision number $n$ at fixed $\lambda=0.05$ for $s=0.05$ (red), $s=0.5$ (blue), and $s=0.9$ (green), with $b_1^{(0)}=1$, $b_2^{(0)}=b_3^{(0)}=0$, and $G_0=1$. The witness is evaluated when the $|\rm Det(\mathcal{E}_{S}(n))|> 10^{-10}$. For $s=0.05$, and $s=0.5$ the witness decays monotonically to zero within 
    roughly ten to thirteen steps. For $s=0.9$ the decay is non-monotone, with oscillations of the order of $10^{-10}$ persisting after $n=13$.}
    \label{fig:Ndiv-n-uniform}
\end{figure}

\begin{figure}[!htb]
    \centering
    \includegraphics[width=0.9\linewidth]{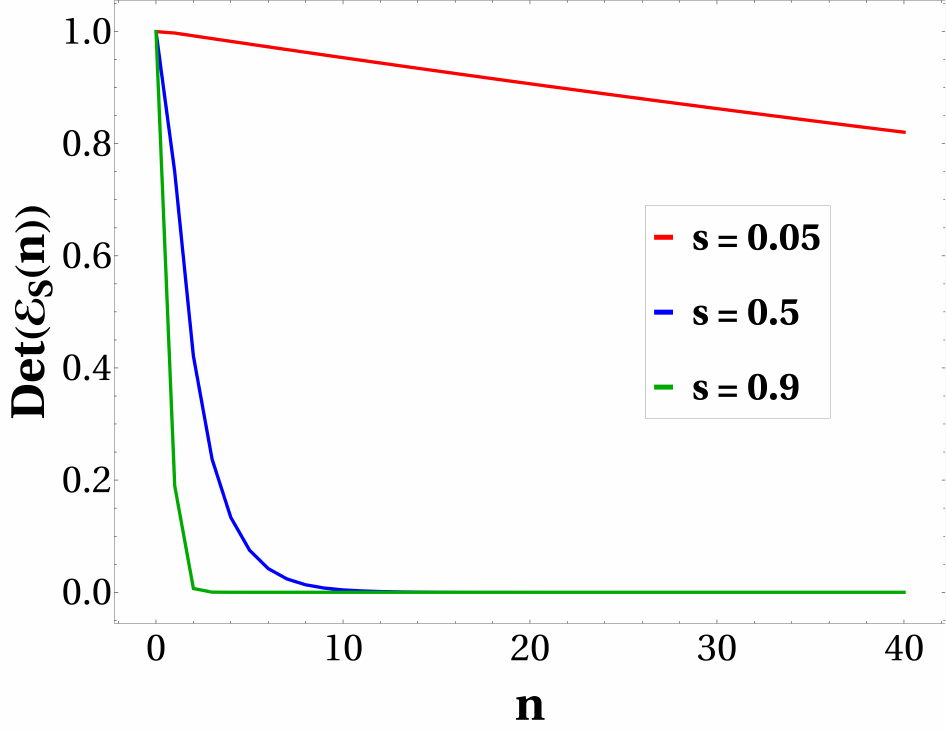}
    \caption{Determinant of the dynamical map $\mathcal{E}_S(n)$ as a function of collision number $n$ at fixed $\lambda=0.05$ for $s=0.05$ (red), $s=0.5$ (blue), and $s=0.9$ (green), with the same initial conditions as Fig.~\ref{fig:Ndiv-n-uniform}. The determinant decays slowly for weak interaction, remaining above $0.8$ at $n=40$ for $s=0.05$ at $n=29$, and falls below $10^{-10}$ within ten steps for $s=0.9$.}
    \label{fig:Det-n-uniform}
\end{figure}

To track how CP-indivisibility is distributed across individual collision steps, and to examine how quickly the dynamical map loses invertibility as correlations accumulate, we fix representative points at $\lambda=0.05$ and $s=0.05, \, 0.5, \, 0.9$  and examine $\mathcal{N}_{\mathrm{div}}(n)$ from Eq.~(\ref{eq:Negative Choi}) and $\rm{Det}(\mathcal{E}_{\rm S}(n))$ for $n$ collisions. The measures are evaluated when $ |{\rm Det}(\mathcal{E}_{\rm S}(n))|>10^{-10}$ to prevent the inverse map from being ill conditioned. Fig.~\ref{fig:Ndiv-n-uniform} shows that for $s=0.05$ and $s=0.5$ the step-wise CP witness decays monotonically to zero within roughly ten to thirteen steps, while for $s=0.9$ the decay is non-monotone with small oscillations of the order of $10^{-10}$ persisting beyond $n=13$ reflecting the building up of strong correlations between S and E. Fig.~\ref{fig:Det-n-uniform} shows the corresponding determinant of $\mathcal{E}_{\rm S}(n)$, which measures the contraction of the Bloch ball and governs the conditioning of the intermediate map $\mathcal{V}_{\rm S}(n, n+1)=\mathcal{E}_{\rm S} (n) \circ \mathcal{E}_{\rm S} (n+1)^{-1}$. The determinant remains above $0.8$ at $n=40$ for $s=0.05$, decays rapidly below $10^{-10}$ at $n=29$ for $s=0.5$, and falls below $10^{-10}$ within ten steps for $s=0.9$, consistent with the strong CP-indivisibility observed at large interaction strength. 


The behavior at small $\lambda$ reveals a physically distinct regime that connects the CP-indivisibility to the finite size of the environmental qubit. For $\lambda\to 0$ the coherent weight $\lambda^n$ in Eq.~(\ref{eq:GD convex map}) becomes negligible after only a few steps, so the dynamics is effectively the semi-group beyond that point, and $\mathcal{N}_\mathrm{div}(n) \to 0$ rapidly. Nevertheless, $\mathcal{N}_\mathrm{div}(n)>0$ persists at every collision, no matter how small $\lambda$ is. Even an arbitrarily strong reservoir cannot reduce the CP-indivisibility to exactly zero for $s>0$. The physical interpretation is a finite-size effect. A macroscopic Markovian reservoir imposes $\lambda=0$ exactly, erasing all correlations instantaneously after each collision and yielding CP-divisible dynamics. The single-qubit environment cannot be a perfect reservoir since it has only one degree of freedom and retains a residual $\lambda\,\chi_{\rm SE}^{(1)}$ for one step (\ref{eq:action of GD on correlation})before the reservoir acts again. This one-step memory, however small, is sufficient to make the intermediate map non-CP. It vanishes only in the exact limit of $s=0$ (no interaction) or $\lambda=0$ (perfect, infinite reservoir), and is otherwise present at every collision, independent of $n$. 

The uniform contraction case studied here provides a tractable setting in which the three dynamical regimes are cleanly separated and the onset of CP-indivisibility can be traced to the residual correlation $\lambda\,\chi_{\rm SE}^{(1)}$ surviving each reset step. The generalized depolarizing channel, by virtue of its isotropic action on all Bloch directions, allows the convex structure of the dynamical map~\eqref{eq:GD convex map} to be exploited, yielding explicit closed-form expressions for all quantities. In the following section we move to the physically more realistic setting of anisotropic contraction, where this factorization is no longer available and the full joint state must be tracked, and examine how the richer correlation structure that results modifies the dynamical regimes identified here.

\subsection{Non-Uniform Contraction of Correlations: Generalised Amplitude Damping Channel}
\label{sec:partial reset:example}

When the environmental reset map contracts the Bloch sphere anisotropically — shrinking different Pauli directions at different rates — the factorization argument used in Sec.~\ref {sec:partial reset:uniform contraction} breaks down. The system--environment correlations built up during each collision can no longer be absorbed into a redefined effective map on $\mathcal{H}_{\rm S}$ alone, and the full joint state $\rho_{\rm SE}$ must be tracked explicitly. To illustrate this regime, we consider the generalized amplitude damping (GAD) channel(\ref{eq:physical GAD}) as the environmental reset map, which contracts the transverse components $\sigma_{1,2}^E$ and the longitudinal component $\sigma_3^E$ at different rates. 

We work out with the equivalent interaction Hamiltonian as in Eq.~\eqref{eq:XX-YY_total}, whereas the initial joint state is taken to be a product state, $\rho_{\rm SE}(0) = \rho_{\rm S}(0)\otimes\rho_{\rm E}(0)$, expressed in the Pauli basis as
\begin{align}
    \rho_{\rm SE}(0) = \frac{1}{4}\Big(\mathbb{1}_{SE}
    &+ a_i^{(0)}\,\sigma_i^S \otimes \mathbb{1}_{\rm E}
    + b_j^{(0)}\,\mathbb{1}_{\rm S} \otimes \sigma_j^E
    \nonumber\\
    &+ c_{ij}^{(0)}\,\sigma_i^S \otimes \sigma_j^E
    \Big),
\end{align}
with the Einstein summation convention implied. Since the initial state is a product state, the initial correlation coefficients satisfy $c_{ij}^{(0)} = a_i^{(0)}b_j^{(0)}$, which will be used as the initial condition for the recurrence relations derived in this section later. Following each S--E collision, the environment undergoes the generalized amplitude damping (GAD) channel. For convenience, we parametrize the GAD channel by putting $e^{-\gamma \tau_1}=\kappa$ in the Eq.~(\ref{eq:physical GAD}), where $\kappa$ acts as the reservoir strength and $G_0 \in [-1,1]$ is the same polarization parameter appearing in Eq.~(\ref{eq:GAD fixed point}), determining the fixed point $\eta_{\rm E}^{\mathrm{GAD}}$.

Note that the polarization of the reset state used in Sec.~\ref{sec:perfect reset:example TI} is related to $G_0$ by $b_3 = -G_0$; the two parameterizations are equivalent and the sign convention of Eq.~(\ref{eq:GAD fixed point}) is used consistently throughout Sec.~\ref{sec:partial reset}. The diagonal entries confirm the anisotropic contraction established in Sec.~\ref{sec:model:realisations}: the transverse components $\sigma_{1,2}^E$ decay with  $\kappa$ (rate $\gamma$), while the longitudinal component $\sigma_3^E$ decays with $\kappa^2 $ (rate $2\gamma$) after each collision. The off-diagonal entry $-G_0(1-\kappa^2)$ encodes the displacement of the fixed point from the maximally mixed state: setting $G_0 = 1$ recovers the standard amplitude damping channel with fixed point $|0\rangle\!\langle 0|$, while $G_0 = -1$ gives the inverted channel with fixed point $|1\rangle\!\langle 1|$. The limits $\kappa \to 1$ and $\kappa \to 0$ correspond to vanishing and complete environmental relaxation, respectively, recovering the identity map and the perfect reset to $\eta_{\rm E}^{\mathrm{GAD}}$ identified in  Sec.~\ref{sec:model:generic}.


\paragraph{Recurrence Relations and Effective Dynamical Map.}
\label{sec:partial reset:recurrence}

One composite step --- S--E collision followed by the GAD channel (Eq.~(\ref{eq:physical GAD}))--- induces a closed set of linear recurrence relations for the system Bloch vector components $\{a_i^{(n)}\}$, the environment Bloch vector component $b_3^{(n)}$, and the correlation coefficients $\{c_{ij}^{(n)}\}$,
\begin{eqnarray}
    a_1^{(n+1)} & = & c\,a_1^{(n)} + s\,c_{32}^{(n)},  \nonumber \\
    a_2^{(n+1)} & = &c\,a_2^{(n)} - s\,c_{31}^{(n)},  \nonumber \\
    a_3^{(n+1)} & = & c^2 a_3^{(n)} + s\big(s\,b_3^{(n)}  + c(c_{21}^{(n)} - c_{12}^{(n)})\big), \nonumber \\
    c_{12}^{(n+1)} & = & \kappa\big(cs(a_3^{(n)}  - b_3^{(n)}) + c^2 c_{12}^{(n)} + s^2 c_{21}^{(n)}\big),  \nonumber \\
    c_{21}^{(n+1)} & = & \kappa\big(cs(b_3^{(n)} - a_3^{(n)}) + c^2 c_{21}^{(n)}  + s^2 c_{12}^{(n)}\big),  \nonumber \\
    c_{31}^{(n+1)} & = & \kappa\big(s\,a_2^{(n)} + c\,c_{31}^{(n)}\big), \nonumber \\
    c_{32}^{(n+1)} & = & \kappa\big(c\,c_{32}^{(n)}  - s\,a_1^{(n)}\big),  \nonumber \\
    b_3^{(n+1)}& = & \kappa^2\big(s(s\,a_3^{(n)}  + c\,c_{12}^{(n)} - c\,c_{21}^{(n)})  + c^2 b_3^{(n)}\big) \nonumber \\
    & & \quad + G_0(\kappa^2 - 1),
    \label{eq:b3 recurrence}
\end{eqnarray}
where $c = \cos(2\alpha\tau)$ and $s = \sin(2\alpha\tau)$ are determined by the interaction strength and collision time. The explicit appearance of the correlation coefficients $c_{ij}^{(n)}$ in the system recurrences confirms that the reduced dynamics of S is not closed and depends on the full joint state at each step, consistent with the partial-reset framework of Sec.~\ref{sec:model}.

The analytical solution of the set of equations \eqref{eq:b3 recurrence} is derived in Appendix~\ref{appendix:Bloch solution Partial reset}, where it is shown that the recurrences decouple into independent transverse and longitudinal sectors governed by the transfer-matrix eigenvalues $\lambda_\pm$ of Eq.~(\ref{eq:eigenvalues}). Using the product-state initial condition $c_{ij}^{(0)} = a_i^{(0)} b_j^{(0)}$, the Bloch vector components after $n$ steps are
\begin{eqnarray}
    a_1^{(n)} & = &  F_1(n)\,a_1^{(0)} + s\,b_2^{(0)}\,F_2(n)\,a_3^{(0)}, \nonumber \\
    a_2^{(n)} & = &  F_1(n)\,a_2^{(0)}  - s\,b_1^{(0)}\,F_2(n)\,a_3^{(0)}, \nonumber \\
    a_3^{(n)} & = & G_{30}(n) - c\,s\,b_2^{(0)}\,F_{32}(n)\,a_1^{(0)} \nonumber\\
    & & \qquad + c\,s\,b_1^{(0)}\,F_{32}(n)\,a_2^{(0)}  + F_{33}(n)\,a_3^{(0)},
    \label{eq:partial reset a3 solution}
\end{eqnarray}
where the functions $F_1(n)$, $F_2(n)$, $G_{30}(n)$, and $F_{3j}(n)$ are defined in Appendix~\ref{appendix:Bloch solution Partial reset}. The effective $n$-step dynamical map $\mathcal{E}_{\rm S}(n)$ defined in Eq.~(\ref{eq:system map}) thereby takes the explicit affine matrix form
\begin{equation}
    \mathcal{E}_{\rm S}(n) =
    {\footnotesize{\begin{pmatrix}
    1 & 0 & 0 & 0 \\
    0 & F_1(n) & 0 & s b_2^{(0)}\! F_2(n) \\
    0 & 0 & F_1(n) & {-s b_1^{(0)} \! F_2(n)} \\
    G_{30}(n) & {-c s  b_2^{(0)}\!F_{32}(n)}
    & c s b_1^{(0)}\!F_{32}(n) & F_{33}(n)
    \end{pmatrix}}},
    \label{eq:Dynamical map Partial Reset}
\end{equation}
acting on the column vector $(1,\,a_1^{(0)},\,a_2^{(0)},\,a_3^{(0)})^T$. We note that $\mathcal{E}_{\rm S}(n)$ is in general not a composition of single-step CPTP maps and it reflects the history dependence generated by the residual correlations $\tilde\chi_{\rm SE}(n)\neq 0$. 

For $\kappa\in(0,1)$ and $|c|<1$, both eigenvalues satisfy $|\lambda_\pm|<1$ and the map admits the unique steady state 
\begin{equation}
    \vec{a}^{\,(\infty)} = (0,\;0,\;-G_0),
    \label{eq:steady state partial reset}
\end{equation}
independently of $\{a_i^{(0)}\}$ and $\{b_i^{(0)}\}$. The system reaches the fixed point $\eta_{\rm E}^{\mathrm{GAD}}$ given in Eq.~(\ref{eq:GAD fixed point}), with the polarization set by $G_0$
alone, consistent with the perfect-reset result of Sec.~\ref{sec:perfect reset:TI}. The explicit forms of the map in the limits $\kappa\to 0$ and $\kappa\to 1$ are given in Appendix~\ref{appendix:limiting cases}, where they are shown to recover, respectively, semi-group-like step-to-step dynamics under maximal reservoir action and pure coherent S--E oscillation in the absence of any reservoir.

\subsection{Dynamical Regimes: Non-Uniform Contraction}
\label{sec:dyn regime non-uniform}

\begin{figure}[!htb]
    \centering
    \includegraphics[width=0.9\linewidth]{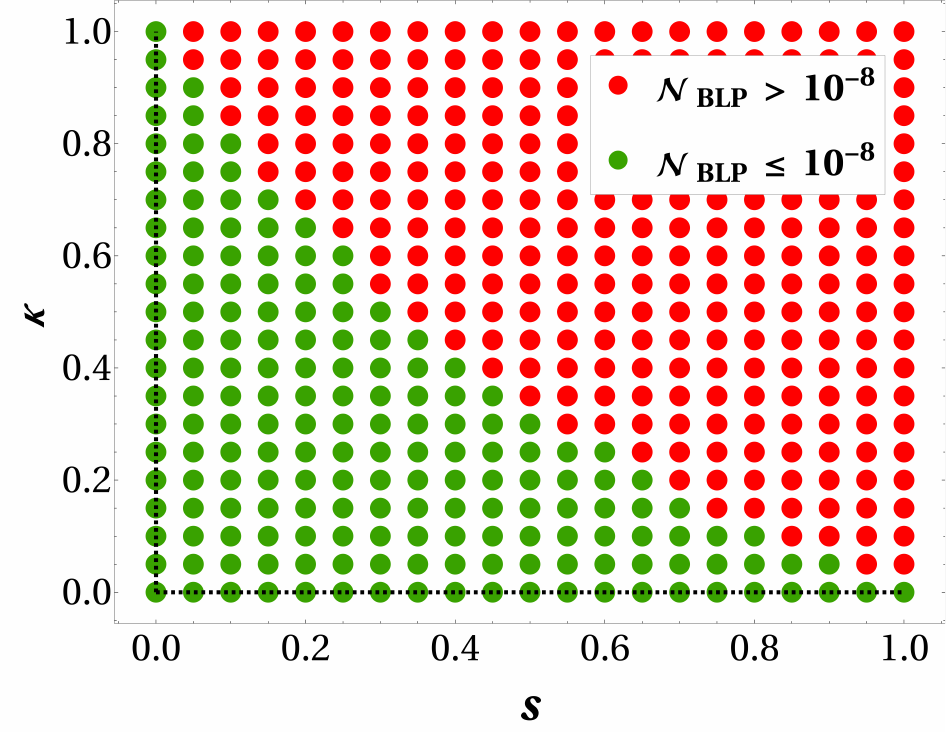}
    \caption{BLP non-Markovianity measure $\mathcal{N}_{\mathrm{BLP}}$ in the $(s,\kappa)$ parameter space, evaluated over $n=100$ collisions with $b_1^{(0)}=1$, $b_2^{(0)}=b_3^{(0)}=0$, and  $G_0=1$. A Markovian region (green) surrounds both axes and is separated from the non-Markovian (red) phase by a boundary $\kappa^*(s)$ that decreases with increasing interaction strength.}
    \label{fig:BLP-measure-non-uniform}
\end{figure}

We now examine the dynamical regimes arising from the map Eq.~\eqref{eq:Dynamical map Partial Reset} for the resonant XX-YY interaction~\eqref{eq:XX-YY_total} with $b_1^{(0)}=1$, $b_2^{(0)}=b_3^{(0)}=0$, and $G_0=1$. Fig.~\ref{fig:BLP-measure-non-uniform} shows $\mathcal{N}_{\mathrm{BLP}}$ over the full $(s,\kappa)$ parameter space, evaluated over $n=100$ collisions. A Markovian region, in which the measure vanishes identically, occupies a neighbourhood of both axes and is bounded above by a phase boundary $\kappa^*(s)$ that moves toward smaller $\kappa$ as $s$ increases. Above this boundary non-Markovian dynamics sets in and $\mathcal{N}_{\mathrm{BLP}}$ grows monotonically with both parameters.

\begin{figure}[!htb]
    \centering
    \includegraphics[width=0.9\linewidth]{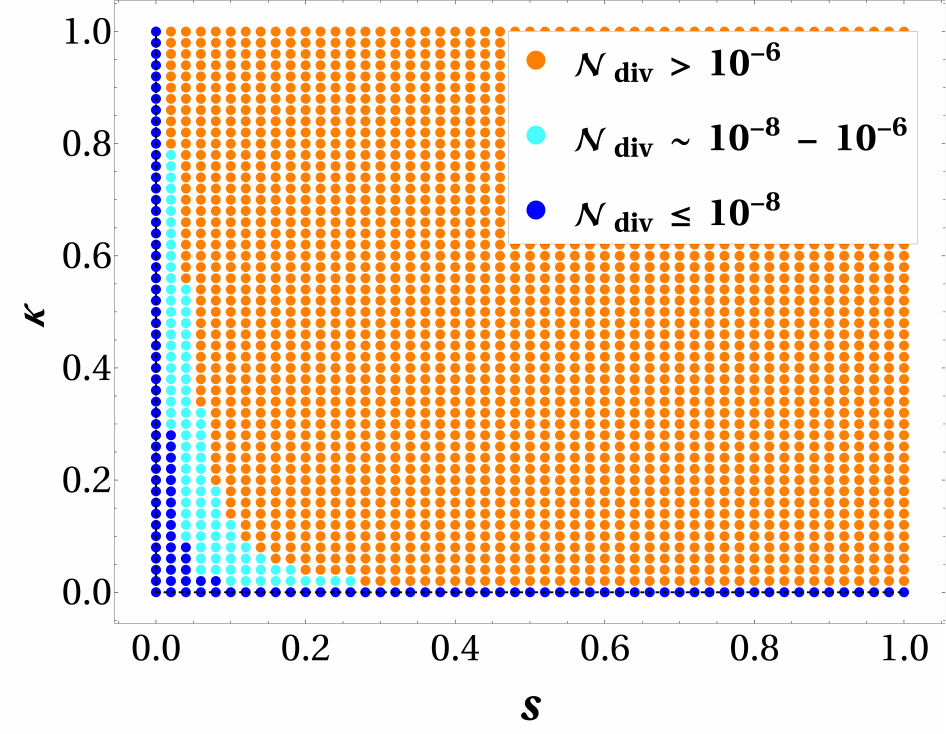}
    \caption{CP-divisibility witness $\mathcal{N}_{\mathrm{div}}$ in the $(s,\kappa)$ parameter space for the same parameters as Fig.~\ref{fig:BLP-measure-non-uniform}. The witness is nonzero throughout the interior $s,\kappa>0$, including in the Markovian region where $\mathcal{N}_{\mathrm{BLP}}=0$, establishing a CP-indivisible yet P-divisible intermediate regime. Unlike the uniform contraction case (Fig.~\ref{fig:Ndiv-uniform}), this intermediate regime is clearly visible on the full-scale plot without requiring any zoom into the small-parameter region.}
    \label{fig:N_div-non-uniform}
\end{figure}

The divisibility measure $\mathcal{N}_{\mathrm{div}}$, shown in Fig.~\ref{fig:N_div-non-uniform}, reveals a richer structure than the BLP measure alone. Throughout most of the interior $s,\kappa>0$ the witness is nonzero, establishing the three-level classification of Table~\ref{tab:divisibility}, namely, CP-divisible along the boundaries, CP-indivisible yet P-divisible within the Markovian region, and non-P-divisible above the BLP phase boundary. While values near the origin (e.g.\ $s,\kappa\approx0.05$) reach $\sim10^{-9}$ and are numerically indistinguishable from zero at machine precision, the intermediate regime is clearly visible at finite $s$ and $\kappa$ on the full-scale plot without requiring any restriction to a small-parameter region, in contrast to the uniform contraction case of Sec.~\ref{sec:partial reset:dyn regime-uniform}. This difference has a structural origin. As $\kappa\to 0$ for any fixed $s>0$ the dynamics approaches the CP-divisible semi group and no transition between regimes can occur in that limit, whereas as $s\to 0$ for fixed $\kappa>0$ a crossover from CP-indivisible to CP-divisible behavior does occur ($\rm orange \to cyan \to blue $) in  Fig.~\ref{fig:N_div-non-uniform}, consistent with the gradual vanishing of the interaction. The anisotropic suppression of the GAD channel --- contracting transverse correlations at rate $\kappa$ and longitudinal correlations at rate
$\kappa^2$ --- combined with the active off-diagonal block $-s\,F_2(n)$ of the map~\eqref{eq:Dynamical map Partial Reset} arising from $b_1^{(0)}=1$, produces a residual correlation structure that is both larger in magnitude and more broadly distributed across the parameter space than in the isotropic GD case, making the CP-indivisibility detectable without having to focus only on the small values of the parameters.

\begin{figure}[!htb]
    \centering
    \includegraphics[width=0.9\linewidth]{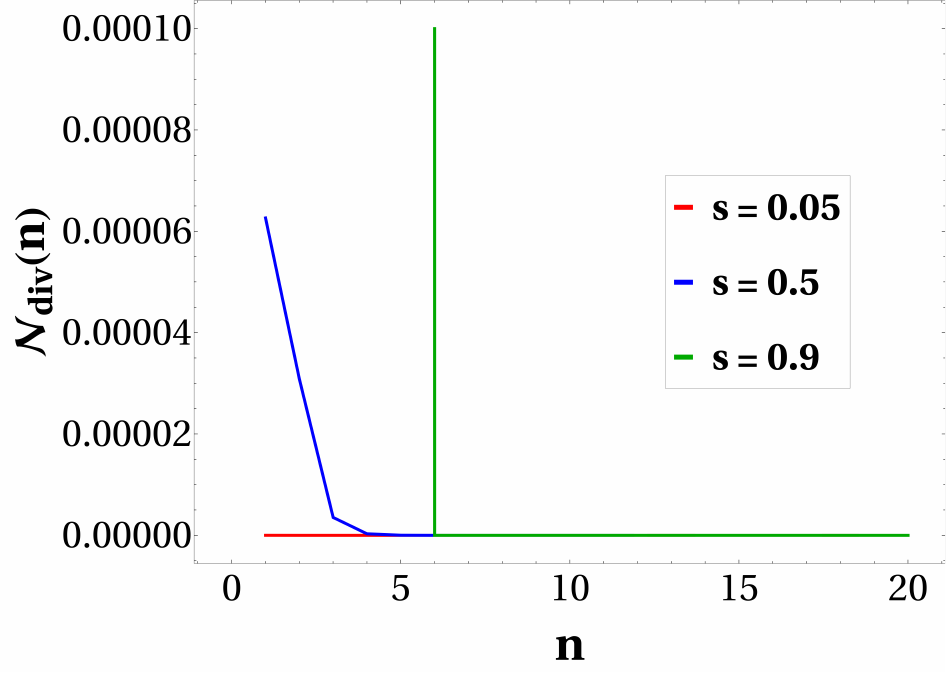}
    \caption{Step-wise CP witness $\mathcal{N}_{\mathrm{div}}(n)$ as a function of collision number $n$ at fixed $\kappa=0.05$ for $s=0.05$ (red), $s=0.5$ (blue), and $s=0.9$ (green), with $b_1^{(0)}=1$, $b_2^{(0)}=b_3^{(0)}=0$, and $G_0=1$. The witness is evaluated only at steps where $\rm |Det(\mathcal{E}_{\rm S} (n))|>10^{-10}$. The  $s=0.05$ line is almost zero (Fig.~\ref{fig:N_div-non-uniform}) from the beginning, and for $s=0.5$, it decays monotonically to zero within five and eleven steps, respectively. For $s=0.9$ it grows to $0.23$  at $n=5$ before the map loses invertibility at $n=6$.}
    \label{fig:Ndiv-n-non-uniform}
\end{figure}

\begin{figure}[!htb]
    \centering
    \includegraphics[width=0.9\linewidth]{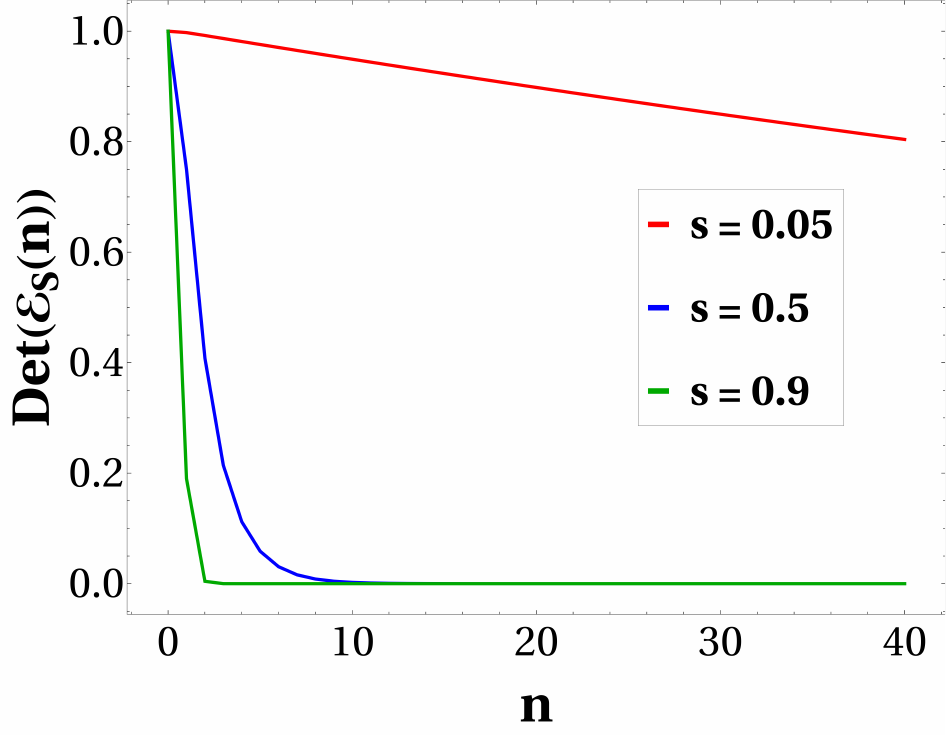}
    \caption{Determinant of the dynamical map $\mathcal{E}_S(n)$ as a function of collision number $n$ at fixed $\kappa=0.05$ for $s=0.05$ (red), $s=0.5$ (blue), and $s=0.9$ (green), with the same initial conditions as Fig.~\ref{fig:Ndiv-n-non-uniform}. All three curves share the same value $0.19$ at $n=1$. The determinant remains above $0.8$ at $n=40$ for $s=0.05$, falls below $10^{-10}$ by $n=29$ for $s=0.5$, and by $n=6$ for $s=0.9$, two steps earlier than in the GD case at the same parameter values, reflecting the accelerated volume contraction induced by the anisotropic GAD channel.}
    \label{fig:Det-n-non-uniform}
\end{figure}

The step-by-step structure is examined by fixing $\kappa=0.05$ and varying the interaction strength. Fig.~\ref{fig:Ndiv-n-non-uniform} shows $\mathcal{N}_{\mathrm{div}}(n)$ as a function of collision number $n$, computed only at steps where $\rm |Det(\mathcal{E}_S(n))|>10^{-10}$ and the intermediate map $\mathcal{V}_S(n,n+1)$ is well defined. For $s=0.05$, the witness opens at $\sim 10^{-8}$, which is almost zero and indicated as a blue color in Fig.~\ref{fig:N_div-non-uniform} as $(0.05,0.05)$. For $s=0.5$, the decay is slower but still monotone, beginning at $\sim 6\times10^{-5}$ and reaching zero by $n=11$. The $s=0.9$ case is qualitatively distinct: rather than decaying, the witness grows from $\sim 10^{-4}$ at $n=1$ to $0.23$ at $n=5$, after which it vanishes abruptly. This is not a physical decay but a hard cutoff imposed by the loss of invertibility of $\mathcal{E}_{\rm S}(n)$ as can be seen from Fig.~\ref{fig:Det-n-non-uniform}, where the determinant for $s=0.9$ falls below $10^{-10}$ by $n=6$, beyond which divisibility cannot be reliably explored.

The determinant data in Fig.~\ref{fig:Det-n-non-uniform} illustrates a key difference from the GD case is that the GAD channel drives the Bloch ball to zero volume considerably faster at the same reservoir parameter value. For $s=0.05$ the determinant remains above $0.8$ throughout ($n=40$), consistent with weak entanglement generation per step. For $s=0.5$ it falls below $10^{-10}$ by $n=29$, and for $s=0.9$ it does so already by $n=6$, compared to ten steps in the uniform case at the same interaction strength. This accelerated contraction originates in the double suppression of the  longitudinal direction at rate $\kappa^2$, which compounds with the transverse suppression at rate $\kappa$ to remove Bloch-ball volume faster than any isotropic channel at the same $\kappa$.

Taken together, the uniform and non-uniform contraction cases establish the central picture of partial-reset dynamics. In both cases the three-level classification of Table~\ref{tab:divisibility} applies, and CP-indivisibility persists for any nonzero interaction strength and reservoir retention. The two cases differ, however, in the weight and distribution of  the intermediate CP-indivisible yet P-divisible regime: under isotropic GD contraction it is confined to a narrow region near the origin and requires a zoom to resolve, while under anisotropic GAD contraction it is broadly distributed and visible at full scale, a direct consequence of the different suppression rates $\kappa$ and $\kappa^2$ acting on the two Bloch sectors. 

In both cases the boundary of the non-Markovian phase is controlled by the competition between the interaction strength $s$ and the reservoir retention parameter ($\lambda \; \rm or\; \kappa $), and CP-divisibility is recovered in the exact limits $s=0$ or vanishing retention. The accelerated loss of map invertibility under GAD contraction, visible in the determinant data of Fig.~\ref{fig:Det-n-non-uniform}, further distinguishes the two cases at the level of individual collision steps. The GD map remains invertible over many more steps at the same parameter values, the GAD map rapidly exhausts the accessible divisibility information, limiting the number of steps over which the witness can be evaluated. These  differences notwithstanding, the qualitative phase structure and the finite-size origin of the irreducible CP-indivisibility are universal features of the partial-reset framework, independent of the specific reservoir model employed.

\section{Conclusion}
\label{sec:conclusion}

In this work, we introduced a minimal collision-model framework in which a system qubit repeatedly interacts with the same environment qubit, with a reservoir-induced CPTP map acting on the environment between successive collisions. Unlike the standard collision model with fresh ancillas at every step, the present construction explicitly retains and controls the system--environment correlations generated during the dynamics, allowing us to investigate their role in the emergence of Markovian and non-Markovian, CP-divisible and indivisible behavior.

We showed that complete suppression of system--environment correlations uniquely requires a full reset operation that drives the environment to a fixed state independently of its history. In the perfect-reset limit, the reduced dynamics recovers a semi group structure with a time-independent GKSL generator, while more general fixed-state resets yield CP-divisible dynamics. For partial resets, the competition between correlation generation and environmental relaxation produces a richer dynamical structure, including CP-divisible, CP-indivisible, Markovian, and non-Markovian regimes.

We further demonstrated that standard noise channels, such as generalised depolarising and amplitude damping processes, naturally realise this reservoir-induced partial-reset framework. Using a resonant excitation-exchange interaction together with a generalised amplitude-damping channel, we explicitly showed that the environmental relaxation strength controls the memory character of the reduced dynamics by suppressing the accumulation of correlations.

Overall, our results provide a physically transparent picture of how effective Markovian behaviour emerges in finite environments through controlled correlation suppression. The framework developed here can be extended to higher-dimensional systems, more general interactions, and thermodynamic settings, and may provide useful insights into memory effects in open quantum systems, quantum thermodynamics, and quantum technologies based on repeated interactions.

\begin{acknowledgments} 
   The authors acknowledge funding support from the National Quantum Mission through the Foundation for Quantum Computing Innovation and I-Hub Quantum Technology Foundation, an initiative of the Department of Science and Technology, Government of India.
\end{acknowledgments}

\appendix

\section{Choi Matrix from the Bloch Dynamical Map}
\label{appendix:choi}

We review the general relation between the affine Bloch representation of a qubit map and its Choi matrix, which is used extensively throughout the paper.

\subsection*{Affine Bloch form}

Any linear map $\mathcal{E}$ on the qubit state space can be written in the ordered Pauli basis $\{\mathbb{1},\sigma_1,\sigma_2,\sigma_3\}$ as an affine map on the Bloch vector,
\begin{equation}
    a_j \;\mapsto\; \sum_k T_{jk}\,a_k + t_j,
    \qquad j=1,2,3,
    \label{eq:app:choi:affine}
\end{equation}
where $T\in\mathbb{R}^{3\times 3}$ is the contraction matrix and
$\vec{t}\in\mathbb{R}^3$ is the translation vector, defined by
\begin{align}
    T_{jk} &= \tfrac{1}{2}\mathrm{Tr}\!\left[\sigma_j\,
              \mathcal{E}[\sigma_k]\right],
    \label{eq:app:choi:Tjk}\\
    t_j    &= \tfrac{1}{2}\mathrm{Tr}\!\left[\sigma_j\,
              \mathcal{E}[\mathbb{1}]\right].
    \label{eq:app:choi:tj}
\end{align}

\subsection*{Choi matrix}

The Choi--Jamio\l{}kowski matrix of $\mathcal{E}$ is defined as \begin{equation}
    \mathcal{C}(\mathcal{E})  = (\mathcal{I}\otimes\mathcal{E}) \!\left[ \ket{\beta_{00}} \! \bra{\beta_{00}} \right],
    \label{eq:app:choi:def}
\end{equation}
where $\ket{\beta_{00}} = (\ket{00}+\ket{11})/\sqrt{2}$ is the maximally entangled Bell state, so that $\mathrm{Tr}[\mathcal{C}]=1$ for any trace-preserving map. Expanding in the Pauli basis and using $\omega_{\mu\nu} \equiv \mathrm{Tr}[\sigma_\mu\sigma_\nu^T]/2 = \mathrm{diag}(1,1,-1,1)_{\mu\nu}$ (the $-1$ is from $\sigma_2^T=-\sigma_2$), the result is 
\begin{equation}
    \mathcal{C}(\mathcal{E})  =\frac{1}{4} \sum_{\mu,\nu=0}^{3} \omega_{\mu\nu} \, \sigma_\mu \otimes \mathcal{E} [\sigma_\nu].
    \label{eq:app:choi:expanded}
\end{equation}
Substituting Eqs.~(\ref{eq:app:choi:Tjk})--(\ref{eq:app:choi:tj}) and evaluating in the computational basis
$\{\ket{00},\ket{01},\ket{10},\ket{11}\}$ gives the explicit $4\times 4$ form
\begin{widetext}
\begin{equation}
    \mathcal{C}(\mathcal{E})
    =\frac{1}{4}
    \begin{pmatrix}
        1+t_3+T_{33}
            & T_{31}-iT_{23}
            & t_1+T_{13}+i(t_2+T_{32})
            & T_{11}-iT_{12}+iT_{21}+T_{22}
        \\[6pt]
        iT_{23}+T_{31}
            & 1+t_3-T_{33}
            & T_{11}+i(T_{12}+T_{21})-T_{22}
            & t_1-T_{13}+i(t_2-T_{32})
        \\[6pt]
        t_1+T_{13}-i(t_2+T_{32})
            & T_{11}-i(T_{12}+T_{21})-T_{22}
            & 1-t_3-T_{33}
            & i(T_{23}+iT_{31})
        \\[6pt]
        T_{11}+i(T_{12}-T_{21})+T_{22}
            & t_1-T_{13}-i(t_2-T_{32})
            & -i(T_{23}-iT_{31})
            & 1-t_3+T_{33}
    \end{pmatrix}\!.
    \label{eq:app:choi:full}
\end{equation}
\end{widetext}
One can verify directly that $\mathrm{Tr}[\mathcal{C}(\mathcal{E})]=1$  and $\mathrm{Tr}_{\rm S} \, \mathcal{C} (\mathcal{E}) = \mathbb{1}_E/2$ for any trace-preserving map, consistent with the normalization convention~\cite{23CHOI1975285,10NielsenChuangQuantum}. The map $\mathcal{E}$ is CP if and only if all eigenvalues of $\mathcal{C}(\mathcal{E})$ are non-negative, which is the condition used to construct the CP-divisibility witness $\mathcal{N}_{\mathrm{div}}$ in Eq.~(\ref{sec:prelim:Divisibility_measure}).

\section{Uniqueness of the Reset Map}
\label{appendix:uniqueness}

We prove that the complete reset map is the unique interaction-independent CPTP map capable of erasing all S-E correlations after a collision. This highlights the fact that Markovian and/or semi-group evolution that is the mainstay of the theory of open quantum systems is more of an exception than the norm.

\subsection*{Erasure condition as a homogeneous system}

After one $\rm{S}$--$\rm{E}$ collision with a unitary $U_{\rm SE}(\tau)$, the joint state in the Pauli basis reads
\begin{align}
    \rho_{\rm{SE}}(1) = \frac{1}{4}\Big(\mathbb{1}_{\rm{SE}}
    &+ a_i^{(1)}\,\sigma_i^{\rm{S}} \otimes \mathbb{1}_{\rm{E}} 
    + b_j^{(1)}\,\mathbb{1}_{\rm{S}}  \otimes \sigma_j^{\rm{E}} 
    \nonumber\\
    &+ c_{ij}^{(1)}\,\sigma_i^{\rm{S}}  \otimes \sigma_j^{\rm{E}} 
    \Big),
    \label{eq:joint-state-one-collision}
\end{align}
with the correlation matrix $\Gamma^{(1)}_{ij} = c^{(1)}_{ij} - a^{(1)}_i b^{(1)}_j$ encoding $\chi_{\rm SE}(1) = \frac{1}{4}\Gamma^{(1)}_{ij}\,\sigma_i^S\otimes\sigma_j^E$. A local CPTP map $\mathfrak{E}$ on $E$ has matrix 
elements $\mathfrak{E}_{jk} = \frac{1}{2} \mathrm{Tr}[\sigma_j^E\,\mathfrak{E}[\sigma_k^E]]$ and translational components $\mathfrak{E}_{j0}$. Under $\mathfrak{E}$, the correlation matrix transforms as
\begin{equation}
    \tilde\Gamma^{(1)}_{ij} 
    = \mathfrak{E}_{jk}\,\Gamma^{(1)}_{ik}, \label{eq:correlation under R}
\end{equation}
with $\mathfrak{E}_{j0}$ dropping out entirely. The erasure condition $\tilde\Gamma^{(1)}_{ij}=0$ therefore becomes the homogeneous system
\begin{equation}
    \mathfrak{E}_{jk}\,\Gamma^{(1)}_{ik} = 0
    \qquad \forall\; i,j \in \{1,2,3\},
    \label{eq:homogeneous}
\end{equation}
which, for each fixed $j$, is a linear system in $\{\mathfrak{E}_{jk}\}_{k=1}^3$ with coefficient matrix $[\Gamma^{(1)}_{ik}]$. A nontrivial solution requires $\rm Det \left([\Gamma^{(1)}]\right) = 0$.

\subsection*{Rank deficiency has no physical realisation}

We show that $\rm Det \left([\Gamma^{(1)}]\right)=0$ cannot arise from any physical post-collision state, so the only solution to Eq.~(\ref{eq:homogeneous}) is $\mathfrak{E}_{jk}=0$ which corresponds to complete reset. Rank deficiency of $[\Gamma^{(1)}]$ requires at least two columns or two rows to be equal; we treat each case in turn.

\paragraph{Equal columns:} Suppose columns $m \neq n$ are equal, $\Gamma^{(1)}_{im} = \Gamma^{(1)}_{in}$ for all $i$, which with $\mu \equiv b^{(1)}_m - b^{(1)}_n$ reads as,
\begin{equation}
    c^{(1)}_{im} = c^{(1)}_{in} + a^{(1)}_i\,\mu
    \qquad \forall\; i.
    \label{eq:col-condition}
\end{equation}
For this state to be physical it must be reachable from Eq.~(\ref{eq:joint-state-one-collision}) by a valid transformation. Since $S$ and all components $b^{(1)}_{n'\neq m}$, $c^{(1)}_{in'\neq m}$ are unchanged, the transformation is necessarily local on $E$, acting as
\begin{equation}
    b^{(1)}_m \;\to\; b^{(1)}_n + \mu, \qquad
    c^{(1)}_{im} \;\to\; c^{(1)}_{in} 
    + a^{(1)}_i\,\mu \quad \forall\; i,
    \label{eq:col-transformation}
\end{equation}
with all other components fixed. Taking $(m,n)=(1,2)$ for concreteness, this local transformation on $E$ is represented in the basis $\{\mathbb{1}_{\rm{E}} ,\sigma^{\rm{E}} _1,\sigma^{\rm{E}} _2,\sigma^{\rm{E}} _3\}$(\ref{eq:app:choi:affine}) by the superoperator
\begin{equation}
    \mathcal{U}_E =
    \begin{pmatrix}
    1 & 0 & 0 & 0\\
    \mu & 0 & 1 & 0\\
    0 & 0 & 1 & 0\\
    0 & 0 & 0 & 1
    \end{pmatrix}.
    \label{eq:UE-matrix}
\end{equation}
The remaining five pairs $(m,n)$ yield superoperators of identical structure with $\mu$ replaced by the corresponding difference $b^{(1)}_m - b^{(1)}_n$, so it suffices to analyse Eq.~(\ref{eq:UE-matrix}).

To assess whether $\mathcal{U}_{\rm E}$ represents a valid physical map, we invoke the Choi--Jamiołkowski isomorphism with $S$ as the reference system — natural since $S$ is a qubit of the same dimension as $E$ and already present in the problem. The map $\mathcal{U}_{\rm E}$ is completely positive if and only if the Choi matrix(\ref{appendix:choi}) is positive semi-definite. Using Eq.~\ref{eq:app:choi:full}, an explicit computation gives
\begin{equation}
    \mathcal{C}(\mathcal{U}_{\rm E}) =
    \begin{pmatrix}
    \tfrac{1}{2} & 0 & \tfrac{\mu}{4} 
        & \tfrac{1-i}{4} \\[4pt]
    0 & 0 & -\tfrac{1-i}{4} & \tfrac{\mu}{4} \\[4pt]
    \tfrac{\mu}{4} & -\tfrac{1+i}{4} & 0 & 0 \\[4pt]
    \tfrac{1+i}{4} & \tfrac{\mu}{4} 
        & 0 & \tfrac{1}{2}
    \end{pmatrix},
    \label{eq:Choi-UE}
\end{equation}
with eigenvalues
\begin{equation}
    {\rm Eig}\big[ \mathcal{C}(\mathcal{U}_{\rm{E}})  \big] \!  \in \! \left\{\!
    -\frac{\sqrt{\mu^2+2}}{4},
    \frac{\sqrt{\mu^2+2}}{4},
    \frac{1}{2}\pm\frac{\sqrt{\mu^2+2}}{4}
    \!\right\}.
    \label{eq:Choi-eigenvalues}
\end{equation}
For every $\mu \in \mathbb{R}$, the minimum eigenvalue $-\frac{1}{4}\sqrt{\mu^2+2} < 0$, so $\mathcal{C}(\mathcal{U}_{\rm E})$ is not positive semi-definite. Hence $\mathcal{U}_{\rm E}$ is not completely positive and the state Eq.~(\ref{eq:col-condition}) is not physical for all $\mu$ and all pairs $(m,n)$.

\paragraph{Equal rows:} Suppose rows $m \neq n$ are equal, $\Gamma^{(1)}_{mi} = \Gamma^{(1)}_{ni}$ for all $i$, which with $\nu \equiv a^{(1)}_m - a^{(1)}_n$ reads as,
\begin{equation}
    c^{(1)}_{mi} = c^{(1)}_{ni} + \nu\,b^{(1)}_i
    \qquad \forall\; i.
    \label{eq:row-condition}
\end{equation}
By the ${\rm{S}}  \leftrightarrow {\rm{E}}$ symmetry of the argument, the transformation required to reach this state is local on $S$, represented by a superoperator $\mathcal{U}_{\rm S}$ of exactly the same structural form as Eq.~(\ref{eq:UE-matrix}) with $\mu \to \nu$, and with $E$ now serving as the reference system in the Choi--Jamiołkowski construction. The Choi matrix is identical in form to Eq.~(\ref{eq:Choi-UE}) with $\mu \to \nu$, giving
\begin{equation}
    {\rm Eig} \big[ \mathcal{C}(\mathcal{U}_{\rm{S}}) \big]_{\min} = 
    -\frac{\sqrt{\nu^2+2}}{4} < 0
    \qquad \forall\; \nu \in \mathbb{R}.
\end{equation}
Hence $\mathcal{U}_{\rm S}$ is not completely positive and the state Eq.~(\ref{eq:row-condition}) is not physical either.

Since every instance of rank deficiency of $[\Gamma^{(1)}]$ leads to an non-physical post-collision state, $\rm Det \left([\Gamma^{(1)}]\right)$ cannot vanish for any physically realizable collision, and the unique solution to Eq.~(\ref{eq:homogeneous}) is
\begin{equation}
    \mathfrak{E}_{jk} = 0 
    \qquad \forall\; j,k \in \{1,2,3\}.
    \label{eq:E-vanishes}
\end{equation}

\subsection*{The reset map is the unique CPTP solution}

Any qubit CPTP map admits the affine decomposition while acting on a state,$\rho_{\rm{E}}=\frac{1}{2}(\mathbb{1}_{\rm{E}}+b_k \sigma^{\rm{E}}_k)$ in Pauli Basis as
\begin{equation}
    \mathfrak{E}[\rho_{\rm E}] = \frac{1}{2}\!\left(
    \mathbb{1}_E 
    + \bigl(\mathfrak{E}_{j0}+\mathfrak{E}_{jk}\,b_k 
    \bigr)\sigma_j^E
    \right).
\end{equation}
Substituting Eq.~(\ref{eq:E-vanishes}), the contraction term vanishes and the output becomes state-independent:
\begin{equation}
    \mathfrak{E}[\rho_{\rm E}] = \eta_{\rm E} 
    = \frac{1}{2}\!\left(
    \mathbb{1}_E + \mathfrak{E}_{j0}\,\sigma_j^E
    \right)
    \qquad \forall\; \rho_{\rm E}.
    \label{eq:reset2}
\end{equation}
This is the complete reset to the fixed state $\eta_{\rm E}$, independent of the input. The target state is parametrized by $\{\mathfrak{E}_{j0}\}$, subject to the Bloch-ball constraint $\sum_j \mathfrak{E}_{j0}^2 \leq 1$. Since every CPTP map with vanishing contraction necessarily takes this form, the reset map is the unique  solution. 

\section{Single-Step Dynamical Map for the XX-YY Interaction}
\label{appendix:single-step map}

We construct the single-step dynamical map corresponding to the interaction Hamiltonian in Eq.~\eqref{eq:XX-YY_total} and a general environmental qubit state written in the Pauli basis $\{\mathbb{1}_E,\sigma^E_j\}_{j=1}^3$,
\begin{equation}
    \rho_{\rm E} = \tfrac{1}{2}\!\left(\mathbb{1}_E + b_j\,\sigma^E_j\right).
    \label{eq:app:general env state}
\end{equation}
Using the definition~\eqref{eq:single step map def TI} and using the results in Appendix.~\ref{appendix:choi}, the components of the single-step map are
\begin{equation}
    \bigl(\Phi^\tau_{\rho_{\rm E}}\bigr)_{jk} \!\!\! = \!\! \tfrac{1}{2}\mathrm{Tr}\!\left[
        \sigma^S_j\,\Phi^\tau_{\rho_{\rm E}}[\sigma^S_k]
      \right],
    \bigl(\Phi^\tau_{\rho_{\rm E}}\bigr)_{j0} \!\!\! = \!\! \tfrac{1}{2}\mathrm{Tr}\!\left[
        \sigma^S_j\,\Phi^\tau_{\rho_{\rm E}}[\mathbb{1}_S]
      \right],
    \label{eq:app:HS components}
\end{equation}
for $j,k\in\{1,2,3\}$. In the ordered basis $\{\mathbb{1}_S,\sigma^S_1,\sigma^S_2,\sigma^S_3\}$ the matrix form is
\begin{equation}
    \Phi^\tau_{\rho_{\rm E}}
    =
    \begin{pmatrix}
        1          & 0           & 0          & 0   \\
        0          & c           & 0          & b_2\,s  \\
        0          & 0           & c          & -b_1\,s \\
        b_3\,s^2   & -b_2\,c\,s  & b_1\,c\,s  & c^2
    \end{pmatrix},
    \label{eq:app:single-step map}
\end{equation}
where $c=\cos(2\alpha\tau)$ and $s=\sin(2\alpha\tau)$. The map depends on the environmental state only through its Bloch-vector components $(b_1,b_2,b_3)$: the transverse components $b_{1,2}$ mix the transverse and longitudinal sectors, while $b_3$ contributes only to the longitudinal shift.

\section{Derivation of the Dynamical Map for Perfect Reset to a Fixed State}
\label{appendix:perfect reset derivation fixed pt}

Specialising Eq.~\eqref{eq:app:single-step map} to the initial environment state
$\rho_{\rm E}(0)=\tfrac{1}{2}(\mathbb{1}_E+b^{(0)}_j\sigma^E_j)$ gives
\begin{equation}
\Phi^\tau_{\rho_{\rm E}(0)} =
\begin{pmatrix}
1                   & 0                    & 0                   & 0    \\
0                   & c                    & 0                   & b^{(0)}_2\,s  \\
0                   & 0                    & c                   & -b^{(0)}_1\,s \\
b^{(0)}_3\,s^2      & -b^{(0)}_2\,c\,s     & b^{(0)}_1\,c\,s     & c^2
\end{pmatrix}.
\label{eq:app:Phi0-fixed}
\end{equation}
After the first collision, the environment is reset at every subsequent step to the fixed longitudinal state $\eta_{\rm E}=\tfrac{1}{2}(\mathbb{1}_E-G_0\sigma^E_3)$, $G_0\in[-1,1]$.
Specialising~\eqref{eq:app:single-step map} to $(b_1,b_2,b_3)=(0,0,-G_0)$ gives
\begin{equation}
\Phi^\tau_{\eta_{\rm E}} =
\begin{pmatrix}
1           & 0  & 0  & 0   \\
0           & c  & 0  & 0   \\
0           & 0  & c  & 0   \\
-G_0\,s^2   & 0  & 0  & c^2
\end{pmatrix}.
\label{eq:app:PhiEta-fixed}
\end{equation}
Because the reset state is purely longitudinal, $\Phi^\tau_{\eta_{\rm E}}$ is block-diagonal and decouples the transverse and longitudinal sectors. We evaluate $\mathcal{E}_S(n)=(\Phi^\tau_{\eta_{\rm E}})^{n-1}\circ\Phi^\tau_{\rho_{\rm E}(0)}$ block by block.

\paragraph{Transverse block.}
The transverse $2\times2$ block of $\Phi^\tau_{\eta_{\rm E}}$ is $c\,\mathbb{1}_2$, so
\begin{equation}
    \bigl(c\,\mathbb{1}_2\bigr)^{n-1} = c^{n-1}\,\mathbb{1}_2.
    \label{eq:app:transverse-fixed}
\end{equation}

\paragraph{Longitudinal block.}
A straightforward induction on the lower-right $2\times2$ block gives
\begin{eqnarray}
\begin{pmatrix}
1 & 0 \\ -G_0\,s^2 & c^2
\end{pmatrix}^{\!\!n-1} \!\!\!\!\! \!
 & = & \!\!\!
\begin{pmatrix}
1 & 0 \\
-G_0\,s^2\displaystyle\sum_{j=0}^{n-2}c^{2j} & c^{2(n-1)}
\end{pmatrix} \nonumber  \\
&= & \!\!\!
\begin{pmatrix}
1 & 0 \\
-G_0(1-c^{2(n-1)}) & c^{2(n-1)}
\end{pmatrix}\!\!, \quad
\label{eq:app:longit-fixed}
\end{eqnarray}
where we used $s^2\sum_{j=0}^{n-2}c^{2j}=1-c^{2(n-1)}$.

\paragraph{Full $n$-step map.}
Composing $(\Phi^\tau_{\eta_{\rm E}})^{n-1}$ with $\Phi^\tau_{\rho_{\rm E}(0)}$ yields
\begin{eqnarray}
\mathcal{E}_S(n) \!\!\! & = & 
\begin{pmatrix}
1 & 0 & 0 & 0 \\
0 & c^{n-1} & 0 & 0 \\
0 & 0 & c^{n-1} & 0 \\
-G_0(1-c^{2(n-1)}) & 0 & 0 & c^{2(n-1)}
\end{pmatrix} \nonumber\\
& & \; \times \begin{pmatrix}
1                & 0                 & 0                & 0    \\
0                & c                 & 0                & b^{(0)}_2\,s  \\
0                & 0                 & c                & -b^{(0)}_1\,s \\
b^{(0)}_3\,s^2   & -b^{(0)}_2\,c\,s  & b^{(0)}_1\,c\,s  & c^2
\end{pmatrix} \nonumber\\
& = & \!\!\!
{\footnotesize{\begin{pmatrix}
1                & 0                      & 0                     & 0         \\
0                & c^n                    & 0                     & b^{(0)}_2 \!\! s c^{n-1}  \\
0                & 0                      & c^n                   & -b^{(0)}_1 \!\!  s c^{n-1} \\
f_{\mathrm{TI}}(n) & -b^{(0)}_2 \!\! s c^{2n-1} & b^{(0)}_1 \!\! s c^{2n-1} & c^{2n}
\end{pmatrix}}} \!, \quad
\label{eq:app:ES-fixed}
\end{eqnarray}
where
\begin{align}
f_{\mathrm{TI}}(n)
= b^{(0)}_3\,s^2\,c^{2(n-1)} - G_0\!\left(1-c^{2(n-1)}\right).
\label{eq:app:fTI}
\end{align}
When $\rho_{\rm E}(0)=\eta_{\rm E}$, i.e.\ $b^{(0)}_3=-G_0$ and $b^{(0)}_1=b^{(0)}_2=0$, the off-diagonal transverse entries vanish and $f_{\mathrm{TI}}(n)$ simplifies to $-G_0(1-c^{2n})$, recovering the semigroup map~\eqref{eq:dynamical map TI} exactly.

\section{Derivation of the Dynamical Map for Perfect Reset to a Varying State}
\label{appendix:perfect reset derivation varying pt}

We now allow the reset state to vary at each step. The environment is still reset to a purely longitudinal state after every collision, but with a step-dependent polarisation
$\eta_{\rm E}^{(n)}=\tfrac{1}{2}(\mathbb{1}_E+G_0^{(n)}\sigma^E_3)$, $G_0^{(n)}\in[-1,1]$. Specialising~\eqref{eq:app:single-step map} to $(b_1,b_2,b_3)=(0,0,G_0^{(n)})$ gives
\begin{equation}
\Phi^\tau_{\eta_{\rm E}^{(n)}} =
\begin{pmatrix}
1              & 0  & 0  & 0   \\
0              & c  & 0  & 0   \\
0              & 0  & c  & 0   \\
G_0^{(n)}\,s^2 & 0  & 0  & c^2
\end{pmatrix}.
\label{eq:app:PhiEta-varying}
\end{equation}
The first collision uses $\Phi^\tau_{\rho_{\rm E}(0)}$ from~\eqref{eq:app:Phi0-fixed} with the general initial state, and all subsequent collisions use maps of the form~\eqref{eq:app:PhiEta-varying}. The $n$-step map is therefore 
$\mathcal{E}_S(n)
= \Phi^\tau_{\eta_{\rm E}^{(n-1)}}\circ\cdots\circ
  \Phi^\tau_{\eta_{\rm E}^{(1)}}\circ\Phi^\tau_{\rho_{\rm E}(0)}$.

\paragraph{Transverse block.}
Each reset map $\Phi^\tau_{\eta_{\rm E}^{(k)}}$ contributes $c\,\mathbb{1}_2$ to the transverse sector, so the $(n-1)$-fold product gives $c^{n-1}\mathbb{1}_2$, identical to~\eqref{eq:app:transverse-fixed}.

\paragraph{Longitudinal block:} The ordered product of the longitudinal blocks of the $n-1$ reset maps is
\begin{equation}
\prod_{k=1}^{n-1}
\begin{pmatrix}
1 & 0 \\ G_0^{(k)}\!s^2 & c^2
\end{pmatrix}
\!\! = \!\! \begin{pmatrix}
1 & 0 \\[4pt]
s^2 \!\! \displaystyle\sum_{m=0}^{n-2} \! \! G_0^{(n-1-m)}\!c^{2m} & c^{2(n-1)}
\end{pmatrix},
\label{eq:app:longit-varying}
\end{equation}
which follows by induction. We see that at step $k$ the polarization $G_0^{(k)}$ contributes  $G_0^{(k)}\,s^2\,c^{2(n-1-k)}$ to the $(3,0)$ entry.

\paragraph{Full $n$-step map:} Composing~\eqref{eq:app:longit-varying} with $\Phi^\tau_{\rho_{\rm E}(0)}$ from Eq.~\eqref{eq:app:Phi0-fixed} gives
\begin{equation}
\mathcal{E}_{\rm S}(n) \!\! = \!\! 
{\footnotesize{\begin{pmatrix}
1                & 0                      & 0                     & 0         \\
0                & c^n                    & 0                     & b^{(0)}_2\! s c^{n-1}  \\
0                & 0                      & c^n                   & -b^{(0)}_1\! sc^{n-1} \\
f_{\mathrm{TD}}(n) & -b^{(0)}_2\! s c^{2n-1} & b^{(0)}_1\! s c^{2n-1} & c^{2n}
\end{pmatrix}}}\!,
\label{eq:app:ES-varying}
\end{equation}
where
\begin{equation}
f_{\mathrm{TD}}(n)
= b^{(0)}_3\,s^2\,c^{2(n-1)}
  + s^2\sum_{m=0}^{n-2}G_0^{(n-1-m)}\,c^{2m}.
\label{eq:app:fTD}
\end{equation}
When $b^{(0)}_1=b^{(0)}_2=0$ this reduces to the diagonal map given in Eq.~\eqref{eq:dynamical map TD} of the main text. Setting $G_0^{(n)}=-G_0$ for all $n$ recovers $f_{\mathrm{TI}}(n)$ from~\eqref{eq:app:fTI}, confirming consistency with Appendix~\ref{appendix:perfect reset derivation fixed pt}.
\section{Solution of the Recurrence Relations}
\label{appendix:Bloch solution Partial reset}

We solve the system of equations in (\ref{eq:b3 recurrence}) by exploiting their sector structure. 

\subsection*{Sector decomposition}

The recurrences decouple into a \textit{transverse sector} $\{a_1,c_{32}\}$, $\{a_2,c_{31}\}$ and a \textit{longitudinal sector} $\{a_3,c_{12},c_{21},b_3\}$. The transverse pairs are each governed by the $2\times 2$ transfer matrix
\begin{equation}
    M = \begin{pmatrix} c & s \\ -\kappa s & \kappa c
    \end{pmatrix},
    \label{eq:transfer matrix}
\end{equation}
with eigenvalues
\begin{equation}
    \lambda_\pm = \frac{c(1+\kappa)}{2}
    \pm\frac{\sqrt{d}}{2},
    \qquad d = c^2(1+\kappa)^2-4\kappa,
    \label{eq:eigenvalues}
\end{equation}
satisfying $|\lambda_\pm|\leq 1$ and $\lambda_+\lambda_-=\kappa$ for $\kappa\leq 1$, $|c|\leq 1$. The longitudinal sector is governed by eigenvalues $\{\kappa,\lambda_+^2,\lambda_-^2\}$, where $c_{12}^{(n)}+c_{21}^{(n)}$ decouples as an independent
mode associated with $\kappa$, reducing the effective dimension to three. Both sectors converge to a fixed point when $|\lambda_\pm|<1$.

\subsection*{Steady state and general solution}

Imposing stationarity gives $\vec{a}^{(\infty)}=(0,0,-G_0)$, consistent with $\eta_{\rm E}^{\mathrm{GAD}}$. Defining $x_n=a_3^{(n)}+G_0$, the general solutions are
\begin{align}
    a_j^{(n)} &= A_j\lambda_+^n+B_j\lambda_-^n,
    \quad j=1,2,
    \label{eq:transverse solution}\\
    a_3^{(n)} &= C_1\kappa^n+C_2\lambda_+^{2n}
    +C_3\lambda_-^{2n}-G_0,
    \label{eq:longitudinal solution}
\end{align}
with transverse coefficients
\begin{equation}
    A_j = \frac{a_j^{(1)}-a_j^{(0)}\lambda_-}
    {\lambda_+-\lambda_-},\qquad
    B_j = \frac{-a_j^{(1)}+a_j^{(0)}\lambda_+}
    {\lambda_+-\lambda_-},
    \label{eq:transverse coefficients}
\end{equation}
and longitudinal coefficients ($i\neq j\neq k$)
\begin{equation}
    C_i = \frac{x_0\mu_j\mu_k
    -x_1(\mu_j+\mu_k)+x_2}
    {(\mu_i-\mu_j)(\mu_i-\mu_k)},
    \label{eq:longitudinal coefficients}
\end{equation}
where $\{\mu_1,\mu_2,\mu_3\}
=\{\kappa,\lambda_+^2,\lambda_-^2\}$ and
\begin{align}
    x_0 &= a_3^{(0)}+G_0,
    \label{eq:x0}\\
    x_1 &= c^2 a_3^{(0)}+s^2 b_3^{(0)}
    -cs\bigl(c_{12}^{(0)}-c_{21}^{(0)}\bigr)+G_0,
    \label{eq:x1}\\
    x_2 &= (c^2-\kappa s^2)^2 a_3^{(0)}
    +c^2 s^2(1+\kappa)^2 b_3^{(0)}\nonumber\\
    &\quad-(1+\kappa)cs(c^2-\kappa s^2)
    \bigl(c_{12}^{(0)}-c_{21}^{(0)}\bigr)\nonumber\\
    &\quad+G_0\bigl(1+(\kappa^2-1)s^2\bigr).
    \label{eq:x2}
\end{align}

\subsection*{Explicit functions for the dynamical map}

Substituting the product-state condition $c_{ij}^{(0)}=a_i^{(0)}b_j^{(0)}$ and simplifying yields the Bloch-vector solutions  Eq.~(\ref{eq:partial reset a3 solution}) and the dynamical map Eq.~(\ref{eq:Dynamical map Partial Reset}) in terms of
\begin{align}
    F_1(n) &= \frac{(c-\lambda_-)\lambda_+^n
    -(c-\lambda_+)\lambda_-^n}
    {\lambda_+-\lambda_-},
    \label{eq:F1}\\
    F_2(n) &= \frac{\lambda_+^n-\lambda_-^n}
    {\lambda_+-\lambda_-},
    \label{eq:F2}\\
    G_{30}(n) &= s^2 F_{30}(n)\,b_3^{(0)}
    +F_{31}(n)\,G_0-G_0,
    \label{eq:G30}\\
    F_{3j}(n) &= \sum_{i=1}^{3}
    l_i(\vec{u}_j)\,\mu_i^n,
    \label{eq:F3j}
\end{align}
where $$l_i(\vec{u})=\frac{u_1\mu_j\mu_k -u_2(\mu_j+\mu_k)+u_3} {(\mu_i-\mu_j)(\mu_i-\mu_k)}\;,\;i\neq j,k,$$ and the reference vectors are
\begin{align}
    \vec{u}_0 &= \bigl(0,\;1,\;c^2(1+\kappa)^2\bigr),
    \label{eq:u0}\\
    \vec{u}_1 &= \bigl(1,\;1,\;1-s^2(1-\kappa^2)\bigr),
    \label{eq:u1}\\
    \vec{u}_2 &= \bigl(0,\;1,\;
    (1+\kappa)(c^2-\kappa s^2)\bigr),
    \label{eq:u2}\\
    \vec{u}_3 &= \bigl(1,\;c^2,\;
    (c^2-\kappa s^2)^2\bigr).
    \label{eq:u3}
\end{align}

\section{Limiting Cases of the Partial-Reset Dynamical Map}
\label{appendix:limiting cases}

We now consider the dynamical map Eq.~(\ref{eq:Dynamical map Partial Reset}) in the two limiting cases $\kappa\to 0$ and $\kappa\to 1$, which connect the partial-reset model to the perfect-reset dynamics of
Sec.~\ref{sec:perfect reset} and to the unitary S--E evolution, respectively.

\subsection*{Maximal reservoir action,
\texorpdfstring{$\kappa\to 0$}{kappa to 0}}

As $\kappa\to 0$, the eigenvalue pair becomes degenerate. Resolving this via the approximation $\{\lambda_+,\lambda_-\}\to\{c,\kappa/c\}$ before taking $\kappa=0$ gives 
\begin{eqnarray*}  
 F_1(n) & \to &  c^n, \\
 F_2(n) & \to & c^{n-1}, \\
 F_{32}(n) & \to &  c^{2(n-1)}, \\
 F_{33}(n) & \to &  c^{2n},  \\
 G_{30}(n) & \to &  (s^2 b_3^{(0)}/c^2)\,c^{2n}-G_0(1-c^{2n}/c^2) \equiv \tilde{f}_{\rm TD}(n), 
 \end{eqnarray*}
 so the dynamical map $\mathcal{E}_{\rm S}(n)\big|_{\kappa=0}$ has the matrix representation,
\begin{equation}
  {\scriptscriptstyle{\begin{pmatrix}
        1 & 0 & 0 & 0 \\
        0 & c^n & 0 &
        s b_2^{(0)}\!c^{n-1} \\
        0 & 0 & c^n &
        -s b_1^{(0)} \! c^{n-1} \\
        \tilde{f}_{\rm TD}(n)
        & -s b_2^{(0)} \! c^{2n-1}
        & s b_1^{(0)} \! c^{2n-1}
        & c^{2n}
    \end{pmatrix}}}.
    \label{eq:kappa0 map}
\end{equation}
The intermediate map from step $n$ to $n+1$ is
\begin{equation}
    \mathcal{V}_{\rm S}(n+1;n)\big|_{\kappa=0}   =
    \begin{pmatrix}
        1 & 0 & 0 & 0 \\
        0 & c & 0 & 0 \\
        0 & 0 & c & 0 \\
        (c^2-1)G_0 & 0 & 0 & c^2
    \end{pmatrix},
    \label{eq:kappa0 intermediate map}
\end{equation}
which is independent of both $n$ and $\{b_i^{(0)}\}$, and is manifestly CPTP for $|c|\leq 1$. The step-to-step dynamics is therefore Markovian and semi group, recovering the structure of Sec.~\ref{sec:perfect reset:TI} at the level of the intermediate map. The appearance of the term $s/c$ in the $n$-step map Eq.~(\ref{eq:kappa0 map}) is a consequence of step ordering and the S--E unitary precedes the reservoir, so the first collision imprints $\rho_{\rm E}(0)$ on the system before any erasure.

\subsection*{No reservoir action,
\texorpdfstring{$\kappa\to 1$}{kappa to 1}}

As $\kappa\to 1$, the external reservoir does not act on the E and the dynamics of S and E reduces to the closed dynamics of two qubits. In this limit, the discriminant of the map matrix is $d\to -4s^2<0$ and the eigenvalues of the map become the unit-modulus pair $\lambda_\pm = e^{\pm i\theta_0}$ with $\theta_0=\cos^{-1}(c)$. The contraction envelope vanishes and the map reduces to
\begin{equation}
    \mathcal{E}_{\rm S}(n)\big|_{\kappa=1}  =
    \begin{pmatrix}
        1 & 0 & 0 & 0 \\
        0 & c & 0 &
        b_2^{(0)} \! s  \\
        0 & 0 & c &
        -b_1^{(0)} \! s  \\
        b_3^{(0)}\! s^2 &
        - b_2^{(0)} \! c s &
       b_1^{(0)} \! c s  &
        c^2
    \end{pmatrix},
    \label{eq:kappa1 map}
\end{equation}
where $c=\cos(2n\alpha\tau) $ and $s = \sin(2n\alpha\tau)$. This corresponds to purely unitary S--E dynamics with no dissipation and no progression to a steady state. The two qubit system undergoes sustained coherent oscillations at frequency $2\alpha$, consistent with $|\lambda_\pm|=1$ and for S, there is no longer any contracting direction and the polarization of S shows periodic recurrences as expected when the S--E system goes through cycles of getting mutually entangled and not-entangled. Note that this limit coincides with the $p_{\rm S}=1$ case of the CPG construction~\cite{44Ciccarello-CM-to-non-Markovian-PhysRevA.87.040103}, in which S interacts persistently with a single environmental qubit without reservoir-induced relaxation.

\bibliographystyle{apsrev4-2} 
\bibliography{references} 

\end{document}